\documentclass[10pt,journal,a4paper,final,oneside,twocolumn]{IEEEtran}

\usepackage{times} 
\usepackage{graphicx}
\usepackage{cite}
\usepackage{citesort}
\usepackage{color}
\usepackage{psfrag}
\usepackage{subfig}
\usepackage{caption}
\usepackage{amssymb}
\usepackage{stfloats}
\usepackage{lmodern}
\usepackage{epsfig}
\usepackage{pifont}
\usepackage{amsmath}

\usepackage{cases}
\usepackage{array}
\usepackage{multicol}
\usepackage{multirow}
\usepackage[T1]{fontenc}
\usepackage{comment}
\usepackage{enumerate}
\usepackage{amsthm}
\usepackage{indentfirst}
\usepackage{setspace}
\usepackage{bm}
\usepackage{float}
\usepackage{cases}
\usepackage[ruled, lined, linesnumbered, commentsnumbered, longend]{algorithm2e}
\SetKwInOut{KwIn}{Input}
\SetKwInOut{KwOut}{Output}
\SetKwProg{Init}{[Initialization]}{}{end}
\SetKwProg{QT}{[Applying the ELA method]}{}{end}
\SetKwProg{DT}{[Outer Layer Iteration]}{}{end}
\SetKwProg{LT}{[Inner Layer Iteration]}{}{end}
\SetKwProg{QB}{[Iteratively implementing the optimization at each RF]}{}{end}
\graphicspath{{figures/}}
\theoremstyle{remark}
\newtheorem{remark}{Remark}
\makeatletter
\renewcommand{\maketag@@@}[1]{\hbox{\m@th\normalsize\normalfont#1}}
\makeatother

\renewenvironment{thebibliography}[1]{
  \begin{oldthebibliography}{#1}
   \setlength{\parskip}{-0.1em}
}
{
  \end{oldthebibliography}
}

\makeatletter
\def\ifundefined{\@ifundefined}
\makeatother 

\begin{document}

\title{UAV-Enabled Joint Sensing, Communication, Powering and Backhaul Transmission in Maritime Monitoring Networks}

\author{Bohan Li,~\textit{Member, IEEE}, Jiahao Liu, Yujun Liang, Qian Li, Haochen Liu, Yaoyuan Zhang, Junsheng Mu,\\ Shahid Mumtaz,~\textit{Senior Member, IEEE}, Sheng Chen,~\textit{Life Fellow, IEEE} %
\thanks{B. Li, J. Liu, Q. Li, Y. Liang and S. Chen are with the Faculty of Information Science and Engineering, Ocean University of China, Qingdao 266100, China (emails: bohan.li, ljh4327, liqian@ouc.edu.cn, 23040031025@stu.ouc.edu.cn, sqc@ecs.soton.ac.uk).} %
\thanks{H. Liu is with School of Electronics and Information, Northwestern Polytechnical University, Xi'an 710019, China (email: haochenliu@nwpu.edu.cn).} %
\thanks{Y. Zhang is with School of Instrumentation Science and Engineering, Harbin Institute
of Technology, Harbin 150001, China (email: zhangyaoyuan@hit.edu.cn).} %
\thanks{J. Mu is with School of Information and Communication Engineering, Beijing University of Posts and
Communications, Beijing 100876, China (email: mu\_junsheng@126.com).} 
\thanks{S. Mumtaz is with Department of Computer Science, Nottingham
Trent University, Nottingham NG1 4FQ, UK (e-mail: Dr.shahid.mumtaz@ieee.org ).}
\vspace*{-5mm}
}

\maketitle

\begin{abstract}
This paper addresses the challenge of energy-constrained maritime monitoring networks by proposing an unmanned aerial vehicle (UAV)-enabled integrated sensing, communication, powering and backhaul transmission scheme with a tailored time-division duplex frame structure. Within each time slot, the UAV sequentially implements sensing, wireless charging and uplink receiving with buoys, and lastly forwards part of collected data to the central ship via backhaul links. Considering the tight coupling among these functions, we jointly optimize time allocation, UAV trajectory, UAV-buoy association, and power scheduling to maximize the performance of data collection, with the practical consideration of sea clutter effects during UAV sensing. A novel optimization framework combining alternating optimization, quadratic transform and augmented first-order Taylor approximation is developed, which demonstrates good convergence behavior and robustness. Simulation results show that under sensing quality-of-service constraint, buoys are able to achieve an average data rate over 22\,bps/Hz using around 2\,mW harvested power per active time slot, validating the scheme's effectiveness for open-sea monitoring. Additionally, it is found that under the influence of sea clutters, the optimal UAV trajectory always keeps a certain distance with buoys to strike a balance between sensing and other multi-functional transmissions.
\end{abstract}


\section{Introduction}\label{S1}

Maritime monitoring is always the paramount approach to exploring the ocean and serves as a critical foundation for marine economic development and environmental protection \cite{liu2015coastal}. However, with the exponential growth in ocean data demand, the current maritime monitoring networks are facing severe challenges in terms of the achievable rates and timeliness of transmission links between monitoring platforms and data center, especially in open sea scenarios. 

To date, the acquisition of monitoring data in open sea is mainly dependent on buoy-to-ship/satellite communication links \cite{wang2018review}. However, subject to the limited communication range and constrained mobility, the ship-assisted data collection in a large sea area is hard and time-consuming. By contrast, satellites employed with the capability of supporting widely-covered communication service has dominated the data gathering method in maritime monitoring networks. For example, the world's largest ocean observation system, Global Ocean Observing System or GOOS for short, and the widely-used Argo networks both utilize buoy-satellite links to obtain pelagic monitoring data \cite{lin2020ocean}. In spite of the extensive coverage, the satellite-aided methods fail to provide robust and high-throughput communication services and only attains kbps-order data rates \cite{andre2015argos}, owing to the fact that buoys equipped with low-cost signal processing and radio frequency (RF) units are struggled with long-distance air-space cross-medium channels. In order to overcome and compensate for the limitations of ship- and satellite-dominated monitoring data collection, the researchers have looked into the unmanned aerial vehicle (UAV)-assisted methods.  

\subsection{Related Works}\label{S1.1}

With the rapid evolution of drone technology, the modern UAV platforms, integrating high-performance baseband processors and RF modules, are expected to improve the efficiency of data collection in complex maritime communication environments \cite{nomikos2022survey}. The authors of \cite{ma2021uav} leveraged the hovering UAV to collect monitoring data from sea surface sensors using non-orthogonal multiple access (NOMA) technique. The work \cite{liu2022energy} proposed an energy-efficient ocean data collection scheme with the aid of UAV, and jointly optimized the transmit power of buoys and UAV trajectory. In the study \cite{nomikos2024improving}, a UAV-swarm aided maritime communication network was proposed to improve connectivity and spectral efficiency based on NOMA. A multi-unmanned devices scheme consisting of UAVs and unmanned surface vehicles (USVs) was designed in \cite{qian2022joint} to enhance the computational capacity of maritime monitoring networks, by optimizing the UAV trajectory, USV's transmit power and computation resource. To remedy the limited coverage of terrestrial base stations (BSs), the authors of \cite{li2020maritime} presented a hybrid satellite-UAV maritime network, in which the UAVs play the role of relay station connected with satellite and terrestrial BSs via backhaul links. However, the above-mentioned papers fail to take the most unique challenges of maritime monitoring networks into consideration.

Specifically, the performance degradation caused by the frequently position variation of monitoring devices, e.g., buoys, under sea waves should be carefully addressed \cite{li2024performance}, especially when the UAV equipped with large number of antennas radiates narrow beams. To this end, the UAV may carry radar device to implement timely sensing \cite{abushakra2021miniaturized} or apply beam-training towards buoys. But the former imposes extra burden on the UAV's overhead, while the latter requires buoys to be equipped with powerful baseband processors and the latency of beam-training also leads to the reduced rates \cite{palacios2017tracking}.  Accordingly, the integrated sensing and communication (ISAC) technique is well-suited to deal with such issues, as the UAV is able to leverage one set of equipment to fulfill the flexible communication and sensing tasks \cite{bayessa2024joint,jing2024isac}, and hence the positions of buoys can be continuously updated, which facilitates the accurate beam matching during data transmissions. To achieve better sensing performance, the effect of sea clutters under irregular sea waves must be considered, which makes the sensing tasks in maritime networks different from that in terrestrial ones \cite{sira2007adaptive}. It is worth noting that the study of UAV-ISAC in maritime networks under the impact of sea clutters has not been explored in open literature.

Another challenge for maritime monitoring networks is the battery life of monitoring devices. The limited battery capacity of buoys severely restricts data offload rates and deployment longevity \cite{wang2023design}. Unlike the cable-powered ground devices, replenishing power for buoys via cables is impractical in open sea. The UAV-assisted wireless charging technique constitutes a potentially feasible solution \cite{xie2021uav} for flexibly and sustainably charging low-cost maritime monitoring devices. The authors of \cite{li2023energy,feng2020joint} proposed a UAV-enabled wireless charging for ground monitoring nodes and maximized the harvested energy by optimizing the UAV trajectory. A UAV-enabled simultaneous wireless information transfer and charging scheme was proposed in \cite{park2023uav}, where the UAV concurrently transmits information and implements charging towards ground nodes based on power splitting policy, after which the ground nodes leverage the harvested power to send data to the UAV. 
Since the RF-based wireless charging exhibits the compatibility to various frequency bands, it can be feasibly incorporated into ISAC using the same frequency resource to form integrated sensing, communication and powering (ISCAP) \cite{li2024integrating}. The authors of \cite{chen2024isac} designed a multi-functional wireless system, where BS adopts ISCAP technique to simultaneously sense target, transmit information and charge ground nodes. A sensing-assisted scheme for robust communication and powering was proposed in \cite{xu2024sensing}, to improve the localization accuracy and power transfer efficiency. It can be implied that ISCAP also holds significant promise in maritime monitoring networks, as it is able to assist UAV in accurately locating monitoring devices as well as establishing stable data transmission and power transferring links with them.  

\subsection{Our Contributions}\label{S1.2}

In this paper, we pioneer the introduction of ISCAP in UAV-enabled maritime monitoring networks, and our novel  contributions are listed as follows.
\begin{itemize}
\item Considering an energy-limited maritime monitoring scenario, we propose a UAV-enabled integrated sensing, communication, powering and backhaul (ISCPB) transmission scheme as well as the tailor-made time-division duplex (TDD)-based frame structure, by which the UAV firstly senses the accurate location of buoys, and then carries out power transfer and uplink transmission with buoys. In the end, to timely process latency-sensitive data, the UAV forwards part of the collected data to the ship via backhaul links. This new multi-functional transmission design improves the rates of data collection under the hostile conditions of buoy energy deficiency in open sea scenario.

\item In the proposed ISCPB scheme, multiple transmission functions are tightly coupled. In particular, the sensing performance highly affects the following uplink transmission and power transferring. Also, the amount of harvested energy at buoys directly determines the uplink rates, which further influences the receivable amount of data at the ship. Therefore, we jointly optimize the time allocation among multiple functions, UAV trajectory, UAV-buoy association and power scheduling, so as to maximize the rates of data collection. To make our proposed scheme robust in maritime scenarios, the sea clutter is practically modeled, and its impact on system performance is carefully analyzed. To the best of our knowledge, this paper is the first in the field of UAV-ISAC that discusses the effect of sea clutters.

\item To solve this challenging optimization problem, we design an optimization method relying on the combination of alternating optimization, quadratic transform and augmented first-order Taylor approximation, which exhibits good convergence behavior and robustness to various buoy monitoring scenarios. We show that during mission period, under the constraint of sensing quality-of-service (QoS), the achievable rates of buoys per time slot is above 22\,bps/Hz with self-sustaining harvested power around 2\,mW per active time slot, which demonstrates that the proposed scheme offers a potential solution to monitoring data collection in open sea scenario.   
\end{itemize} 

\section{System Model}\label{S2}

We consider a practical scenario of data collection in open sea, where {$U$} monitoring buoys over the sea surface are dragged by the underwater anchors to avoid drifting away, and a rotary-wing UAV based in nearby ship is assigned to collect the observation data and provide power supply, from and to the energy-limited buoys, respectively. In order to deal with the buoy fluctuation caused by the sea wave, the UAV carries out the sensing tasks of localization and tracking towards the buoys such that the antenna beams can be accurately aligned during uplink communication and wireless charging. Moreover, considering the demand of real-time data processing and the limited storage on board the UAV, a portion of the collected data are timely offloaded to the ship via wireless backhaul links during mission period. 
 
A squared uniform plane array (UPA) with $R^2$ antennas is mounted on the UAV and parallel to the sea surface. Each buoy is equipped with two separate antennas for transmitting the uplink data and harvesting the energy, respectively. The ship is equipped with a single antenna for receiving the backhaul data.
Let $T$ be the UAV mission period, which is discretized into $N$ time slots (TSs). The length of each TS $T_{\text{t}}\! =\! \frac{T}{N}$ is set to be sufficiently small, during which the location of the UAV remain approximately unchanged.
Since the buoys are usually sparsely distributed in the open sea, we assume that the UAV only serves up to one buoy within each TS for the sake of power efficiency and computational complexity. Let $\alpha_u[n]\! \in\! \left\{0,1\right\}$ be the association indicator between the $u$-th buoy and UAV in the $n$-th TS. If $\alpha_u[n]\!=\! 1$, the UAV executes the hybrid tasks towards the $u$-th buoy within the $n$-th TS, and $\alpha_u[n]\! =\! 0$ otherwise. Let $\mathcal{U}\! =\! \left\{1,\dots ,U\right\}$ denote the set of buoys. We have $\sum_{u\in \mathcal{U}}\alpha_u[n]=1$ for any TS $n$.

\begin{figure}[h]
\vspace*{-3mm}
\centering
\includegraphics[width=0.8\linewidth]{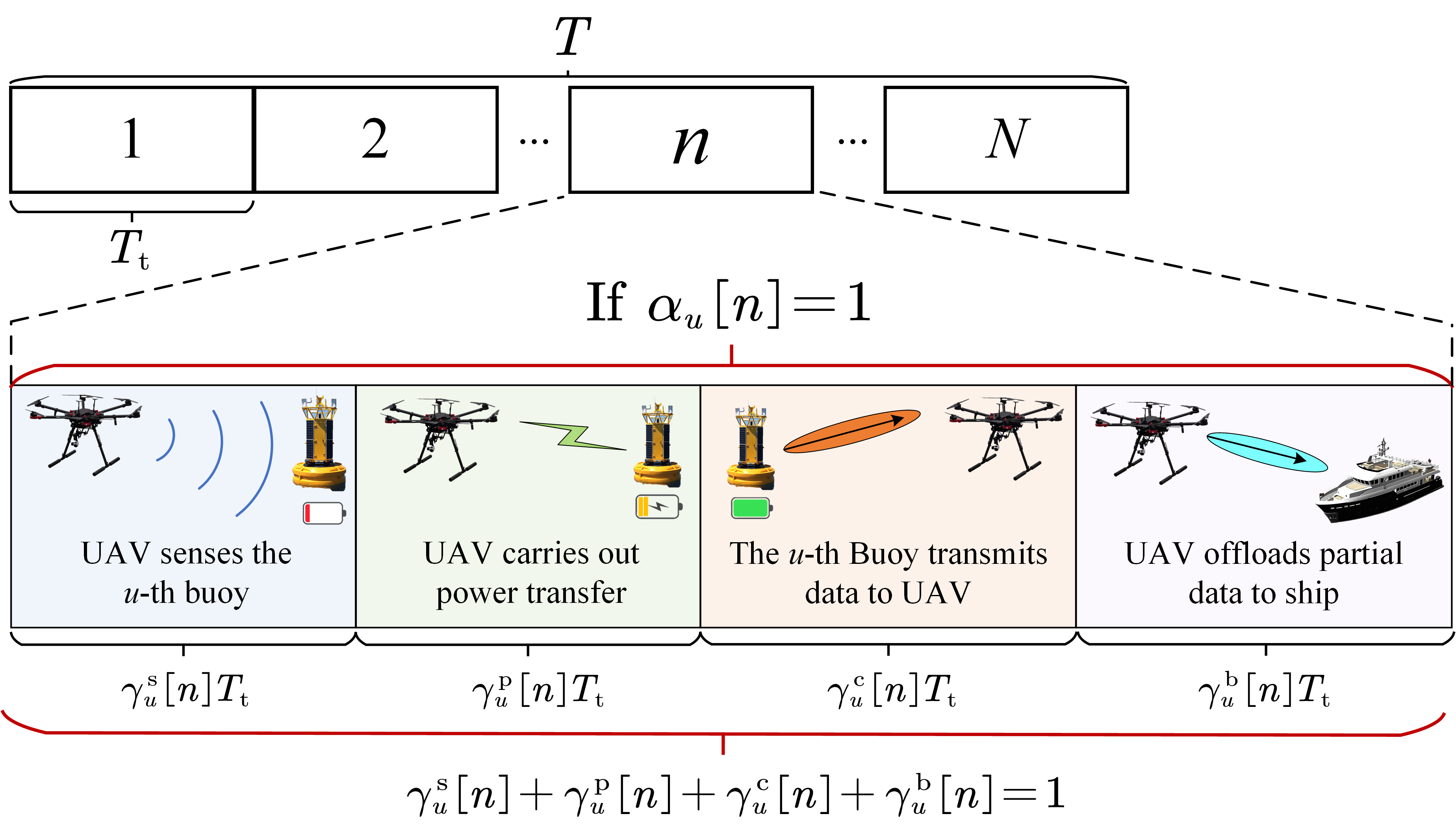}
\vspace*{-2mm}
\caption{\small Frame structure deigned for UAV's ISCPB mission.}
\label{figure-MDDSCC-FS} 
\end{figure}

To avoid the cross-interference among the hybrid tasks of sensing, wireless charging and communication, a TDD-like operation is adopted within every TS of the tailor-made frame structure, which is shown in Fig.~\ref{figure-MDDSCC-FS}. Each TS is composed of four stages. At the first stage, relying on the prior knowledge of buoy's rough coordinates obtained by the GPS, the UAV leverages the sensing signal to precisely localize or track the position of buoy $u$\footnote{At the beginning, with the aid of GPS information, the UAV radiates sensing signal to scan the target area. Once buoy $u$ is detected, the accurate localization is readily realized. In the following TSs related to buoy $u$, i.e., $\left\{n|\alpha_u[n]=1\right\}$, the UAV can simply scan the area around the initial position to achieve tracking.}. After obtaining the accurate position information, the UAV implements the power transfer to and receives the data from buoy $u$, at the second and third stages, respectively. At the fourth stage, the UAV offloads a portion of the data received at the current TS to the ship via wireless backhaul links. The length of these four stages depends on the specific TS, amounting to $\gamma_u^{\text{s}}[n]T_{\text{t}}$, $\gamma_u^{\text{p}}[n]T_{\text{t}}$, $\gamma_u^{\text{c}}[n]T_{\text{t}}$ and $\gamma_u^{\text{b}}[n]T_{\text{t}}$, respectively, with $\gamma_u^{\text{s}}[n]\! +\! \gamma_u^{\text{p}}[n]\! +\! \gamma_u^{\text{c}}[n]\! +\! \gamma_u^{\text{b}}[n]\! =\! 1$, $\forall u,n$.   

\subsection{UAV Sensing}\label{S2.1}

Assume that the UAV flies sufficiently high, such that the channels between the UAV and buoys are largely dominated by line-of-sight (LoS) links. Thus, the sensing channel of the $n$-th TS between the UAV and the $u$-th buoy can be expressed as
\begin{align}\label{eqLoSch} 
  \pmb{H}^{\text{s}}_{u}[n] =& \sqrt{\frac{G_{\mathrm{UAV}}^{\mathrm{tx}} G_{\mathrm{UAV}}^{\mathrm{rx}} \lambda^2 \psi_u}{( 4\pi )^3 d_{u}^{4}[n]}} e^{-j 2\pi \tau_u[n] f_{\text{c}}} \nonumber \\
  & \hspace*{-10mm}\times e^{j 2\pi f^{\mathrm{D}}_u[n] t_0} \pmb{a}_{\mathrm{tx}}\big( \theta_{u}[n],\phi_{u}[n]\big) \pmb{a }_{\mathrm{rx}}^{\rm H}\big( \theta_{u}[n],\phi_{u}[n]\big),\!
\end{align}
where $G_{\mathrm{UAV}}^{\mathrm{tx}}$, $G_{\mathrm{UAV}}^{\mathrm{rx}}$, $\lambda$, $\psi_u$, $\tau_u[n]$, $f_{\text{c}}$, $f^{\mathrm{D}}_u[n]$ and $t_0$ denote the antenna gains of UAV transmitter and receiver, wavelength, radar cross-section (RCS), path delay, central frequency, Doppler frequency and symbol duration, respectively, while $d_{u}[n]$ denotes the distance between the UAV and buoy $u$ at TS $n$, $\theta_{u}[n]$ and $\phi_{u}[n]$ denote the elevation angle of departure/arrival (E-AoD/AoA) and the azimuth angle of departure/arrival (A-AoD/AoA), respectively. Both the UAV's transmit/receive array response vectors, $\pmb{a}_{\mathrm{tx}}\big( \theta_{u}[n],\phi_{u}[n]\big)$/$\pmb{a }_{\mathrm{rx}}\big( \theta_{u}[n],\phi_{u}[n]\big)$ can be expressed as:
\begin{align}\label{eqARV} 
  & \pmb{a}\left(\theta_{u}[n],\phi_{u}[n]\right) = \frac{1}{R}\left[1,\dots,e^{\frac{-j 2\pi (R-1) d}{\lambda}\sin(\theta_{u}[n])\cos(\phi_{u}[n])}\right]^{\text{T}} \nonumber \\
  & \hspace*{18mm}\otimes\left[1,\dots,e^{\frac{-j 2\pi (R-1) d}{\lambda}\sin(\theta_{u}[n])\sin(\phi_{u}[n])}\right]^{\text{T}}\!\! ,\!
\end{align}
where $d$ is the antenna spacing along both the directions of x-axis and y-axis. Assume that the UAV flies at the constant height, and its 3D coordinate within the $n$-th TS is $\pmb{c}_{\text{UAV}}[n]\! =\! [x_{\text{UAV}}[n],y_{\text{UAV}}[n],z_{\text{UAV}}]^{\text{T}}$. The 3D coordinates of buoys are denoted by $\pmb{c}_u[n]\! =\! [x_u[n],y_u[n],z_u[n]]^{\text{T}}, \forall u,n$. Then, given $d_u[n]\! =\! \left\|\pmb{c}_{\text{UAV}}[n]-\pmb{c}_u[n]\right\|$, we have $\sin(\theta_{u}[n])\cos(\phi_{u}[n])\! =\! \frac{x_{\text{UAV}}[n]-x_u[n]}{d_u[n]}$ and $\sin(\theta_{u}[n])\sin(\phi_{u}[n])\! =\! \frac{y_{\text{UAV}}[n]-y_u[n]}{d_u[n]}$. Therefore,  for the sensing channel, $\pmb{a}\left(\theta_{u}[n],\phi_{u}[n]\right)$ can be rewritten as $\pmb{a}\left(\pmb{c}_{\text{UAV}}[n],\pmb{c}_u[n]\right)$.

\begin{figure}[t]
\centering
\includegraphics[width=0.65\linewidth]{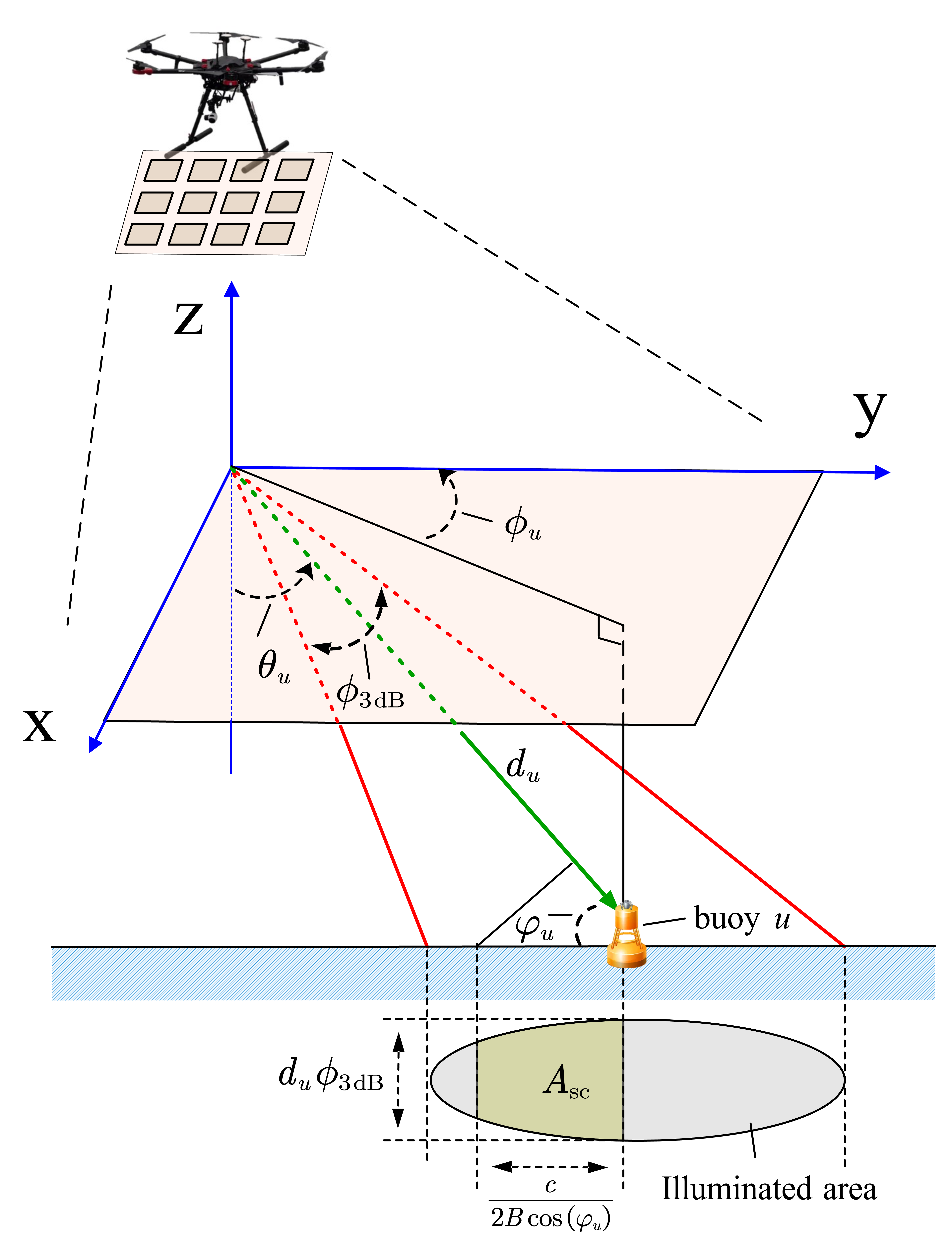}
\vspace*{-3mm}
\caption{\small Influence of sea clutter during UAV sensing.}
\label{figure-MDDSCC-radar} 
\vspace*{-2mm}
\end{figure}

Let $x^{\text{s}}_u[n]$ be the sensing signal at the $n$-th TS with $\mathbb{E}\big\{\left|x^{\text{s}}_u[n]\right|^2\big\}\! =\! 1$. Then, the UAV transmitter emits the signal $\pmb{s}_{\text{s}}[n]\! =\! \sum_{u\in\mathcal{U}}\alpha_u[n]\pmb{f}_u[n]x^{\text{s}}_u[n]$, where $\pmb{f}_u[n]\! \in\! \mathbb{C}^{R^2}$ is the sensing precoder. To process the echo signal reflected by the $u$-th buoy, the sensing combiner $\pmb{w}_u[n]\! \in\! \mathbb{C}^{R^2}$ is applied, and the echo signal after combining is given by
\begin{align}\label{eqCES} 
  r_u[n] =& \pmb{w}_u^{\text{H}}[n] \left(\pmb{H}^{\text{s}}_{u}[n]+\pmb{H}_u^{\text{sc}}[n]\right)^{\text{H}} \pmb{s}_{\text{s}}[n] + \pmb{w}_u^{\text{H}}[n] \pmb{n}_{\text{s}}[n] ,
\end{align}
where $\pmb{n}_{\text{s}}[n]\! \in\! \mathbb{C}^{R}\! \sim\! \mathcal{CN}\left(\pmb{0},N_0\pmb{I}_R\right)$ denotes the additive white Gaussian noise (AWGN). As shown in Fig.~\ref{figure-MDDSCC-radar}, $\pmb{H}_u^{\text{sc}}[n]$ is the interference channel arisen from the sea clutter patch around the $u$-th buoy, which can be expressed as
\begin{align}\label{eqEC} 
  & \pmb{H}_u^{\text{sc}}[n] =  \sqrt{\frac{G_{\mathrm{UAV}}^{\mathrm{tx}} G_{\mathrm{UAV}}^{\mathrm{rx}} \lambda^2 \sigma^{\text{sc}}_u[n] A^{\text{sc}}_u[n]}{( 4\pi )^3 d_{u}^{4}[n]}} e^{-j 2\pi \tau^{\text{sc}}_u[n] f_{\text{c}}} \nonumber \\
  & \times\! e^{j 2\pi f^{\mathrm{sc,D}}_u[n] t_0} \pmb{a}_{\mathrm{tx}}\! \big( \pmb{c}_{\text{UAV}}[n],\pmb{c}_{u}[n]\big) \pmb{a}_{\mathrm{rx}}^{\rm H}\big( \pmb{c}_{\text{UAV}}[n],\pmb{c}_{u}[n]\big),\!
\end{align}
where $\sigma^{\text{sc}}_u[n]$ and $A^{\text{sc}}_u[n]$ denote the backscattering coefficient and the area of sea clutter patch, respectively. The backscattering coefficient can be formed by the widely-used Morchin model \cite{clarke1985airborne}, which is given by
\begin{align}\label{eqMm} 
  \sigma^{\text{sc}}_u[n] =& \underbrace{\frac{4\times 10^{0.6\left(\kappa_{\text{s}}+1\right)-7}\sigma_0[n]\sin\varphi_{u}[n]}{\lambda}}_{\sigma^{\text{sc,1}}_u[n]} \nonumber \\
  & + \underbrace{\Gamma_\text{s}e^{-\tan^2\left(\pi/2-\varphi_{u}[n]\right)\Gamma_\text{s}}}_{\sigma^{\text{sc,2}}_u[n]} ,
\end{align} 
where the integer $0\! \leq\! \kappa_{\text{s}}\! \leq\! 9$ defines the sea state, $\varphi_{\text{u}}[n]$ is the grazing angle, $\Gamma_\text{s}=\cot^2\left(2.44(\kappa_{\text{s}}+1)^{1.08}/57.29\right)$, and $\sigma_0[n]$ is defined by
\begin{equation}\label{eqMm1} 
  \sigma_0[n] = \begin{cases}
    \left(\frac{\varphi_{u}[n]}{\varphi_{0}}\right)^{1.9}, \ &\varphi_{u}[n]\leq\varphi_{0} , \\
    1,\  &\varphi_{u}[n] > \varphi_{0} ,\ 
  \end{cases}
\end{equation}
with $\varphi_{0}\! =\! \text{arcsin}\left(\lambda/(0.1\pi+0.184\pi\kappa_{\text{s}})\right)$.

\begin{figure}[t]
\vspace*{-2mm}
\centering
\includegraphics[width=0.85\linewidth]{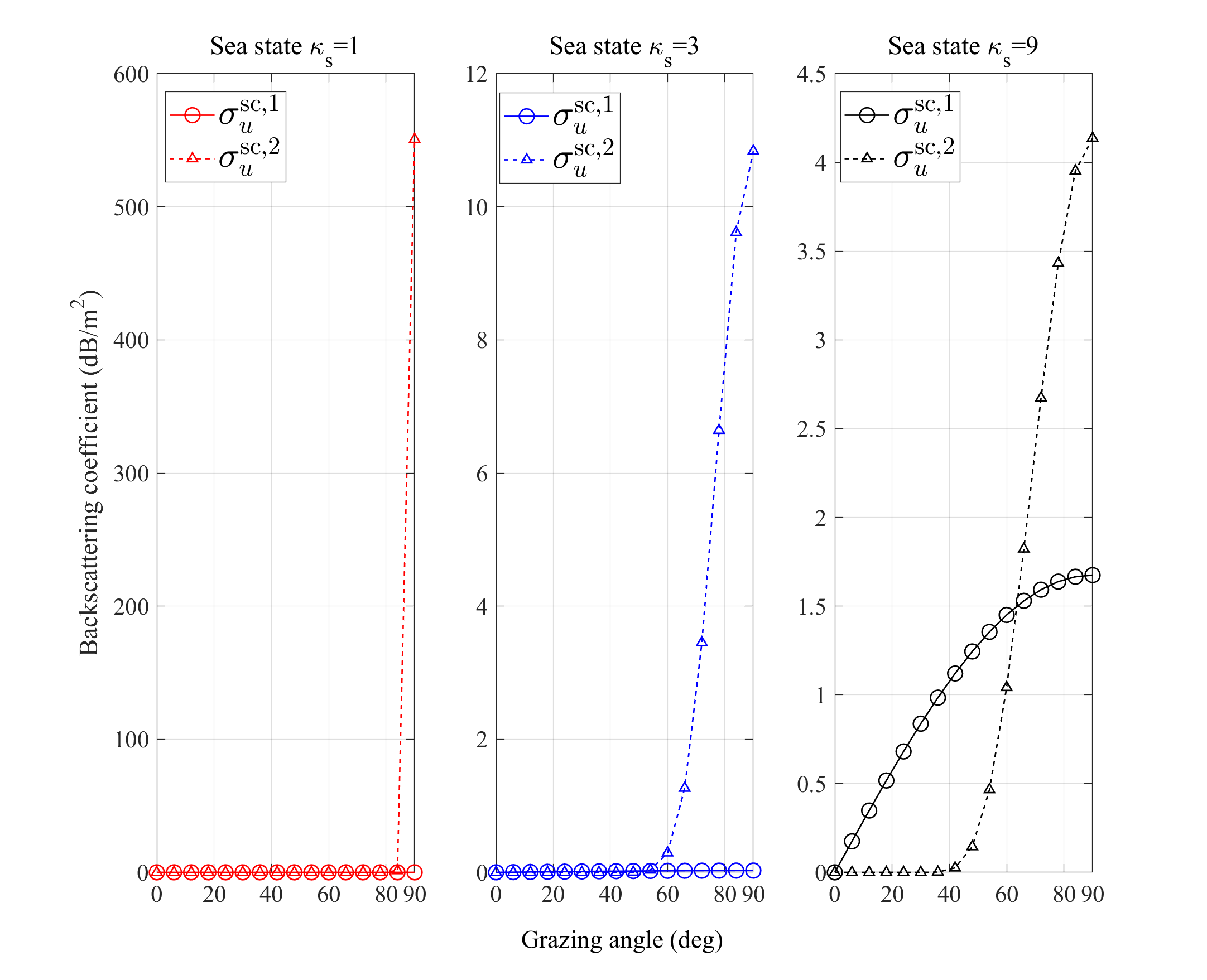}
\vspace*{-3mm}
\caption{\small Contributions of $\sigma^{\text{sc,1}}_u$ and $\sigma^{\text{sc,2}}_u$ to the backscattering coefficient under different sea states.}
\label{figure-MDDSCC-sc} 
\vspace*{-1mm}
\end{figure}

\begin{remark}\label{Rm1}
According to (\ref{eqMm}), the backscattering coefficient is a function of sea state and grazing angle, and consists of two parts. Fig.~\ref{figure-MDDSCC-sc} shows the contributions of $\sigma^{\text{sc,1}}_u$ and $\sigma^{\text{sc,2}}_u$ to $\sigma^{\text{sc}}_u$ under different grazing angles and sea states. It can be seen that unless the sea condition is severe, $\sigma^{\text{sc,2}}_u$ dominates the backscattering coefficient. In the sequel, $\sigma^{\text{sc,1}}_u$ is neglected for analysis convenience.
\end{remark}

Moreover, the area of sea clutter patch can be derived as $A^{\text{sc}}_u[n]\! =\! \frac{c\, d_u[n]\phi_{\text{3dB}}}{2B\cos\left(\varphi_{u}[n]\right)}$, where $c$ and $B$ are light speed and bandwidth, respectively, while $\phi_{\text{3dB}}$ denotes the half-power beamwidth in the azimuth plane, which is given by $\phi_{\text{3dB}}\! =\!\frac{0.886\lambda}{R\cdot d}$ \cite{balanis2015antenna}. Generally speaking, given the formulation of $\sigma^{\text{sc}}_u[n] A^{\text{sc}}_u[n]$, it can be implied that the larger the grazing angle is, the worse sea clutter may occur.

We adopt the sensing mutual information (MI) as the metric \cite{ouyang2023integrated} to evaluate the sensing performance towards buoy $u$ at the $n$-th TS, which can be expressed as
\begin{align}
  & R_u^{\text{s}}[n] = \gamma_u^{\text{s}}[n] \nonumber \\
  & \times\! \log\! \Bigg(\! \! 1+\! \frac{\alpha_u[n]\big|\pmb{w}_u^{\text{H}}[n] \left(\pmb{H}^{\text{s}}_{u}[n]\right)^{\text{H}}\pmb{f}_u[n]\big|^2}{\big|\pmb{w}_u^{\text{H}}[n] \left(\pmb{H}^{\text{sc}}_{u}[n]\right)^{\text{H}}\pmb{f}_u[n]\big|^2\! +\! N_0\left\|\pmb{w}_u[n]\right\|^2}\! \Bigg)\! .\!
\end{align}
Note that the sensing MI has been widely used in ISAC, since it not only has the similar properties and the same measurement unit as communication MI, but also can provide a universal lower bound for the estimation-theoretic metrics, e.g., Cram{\'e}r-Rao bound and mean squared error, regardless of the specific estimator \cite{10540149,10411942,10679234}. Additionally, in our proposed frame structure, the stage of UAV sensing lasts for a while, and it is intuitive that the longer duration of sensing can lead to more accurate results of localization or tracking. Hence the sensing MI is a better metric to take the effect of sensing time into account. 

\subsection{UAV Power Transferring}\label{S2.2}

Each buoy is equipped with an energy receiver (ER), in which the received RF signals are converted into direct current (DC) signals for energy harvesting. Assume that the RF-to-DC energy conversion of the rectifier is linear, and the maximization of the harvested DC power amounts to that of the harvested RF power. As a result, the harvested RF power at the ER of buoy $u$ within the $n$-th TS can be expressed as
\begin{align}\label{eqRH} 
  P_u^{\text{p}}[n] =& \gamma_u^{\text{p}}[n] \xi_u\Big|\big(\breve{{\pmb{h}}}_u^{\text{p}}[n]\big)^{\text{H}} \pmb{v}_u[n]\alpha_u[n]x_u^{\text{p}}[n]\Big|^2 ,
\end{align}
where $0\! <\! \xi_u\! <\! 1$ is the energy conversion efficiency depending on the hardware circuit of buoy $u$, $x_u^{\text{p}}[n]$ with $\mathbb{E}\big\{\left|x^{\text{p}}_u[n]\right|^2\big\}\! =\! 1$ is the power signal transmitted at the $n$-th TS, and $\pmb{v}_u[n]\! \in\! \mathbb{C}^{R^2}$ is the power transfer precoder. 

\begin{figure*}[!b]\setcounter{equation}{15}
\vspace*{-4mm}
\hrulefill
\begin{align}\label{eq:MDDSCC:Rus} 
  R_u^{\text{s}}[n] =& \gamma_u^{\text{s}}[n]\alpha_u[n] \log\left(1+\frac{ G_{\text{gain}}^{\text{s}}\lambda^2 \psi_u R B  \bar{d}_u[n]}{0.886 c\, G_{\text{gain}}^{\text{s}} \lambda^2 \Gamma_{\text{s}}e^{-\Gamma_{\text{s}}\left(\frac{\bar{d}_u[n]}{z_{\text{UAV}}-z_u[n]}\right)^2} d^2_u[n]+N_0RB 4\pi^3 d^4_u[n]\bar{d}_u[n]}\right) ,
\end{align}
\vspace*{-5mm}
\end{figure*} 

It should be mentioned that although buoys are dragged by the underwater anchors, their positions will slightly change within each TS due to sea waves. Specifically, buoy $u$ deviates a little bit from the position that was sensed by the UAV at the beginning of the TS, i.e., $\breve{\pmb{c}}_u[n]\! =\! [x_u[n]\! +\! e_x[n],y_u[n]\! +\! e_y[n],z_u[n]\! +\! e_z[n]]^{\text{T}}$, with $e_x[n], e_y[n],e_z[n]\! \in\! \mathcal{N}(0,\epsilon^2)$. Then the channel between the $u$-th buoy and the UAV, i.e., $\breve{{\pmb{h}}}_u^{\text{p}}[n]\! \in\! \mathbb{C}^{R^2}$, is given by \setcounter{equation}{8}
\begin{align}\label{eq:MDDSCC:hpu} 
  \breve{{\pmb{h}}}_u^{\text{p}}[n] =& \sqrt{\frac{G_{\mathrm{UAV}}^{\mathrm{tx}} G_{\mathrm{u}}^{\mathrm{rx}} \lambda^2}{(4\pi)^2 \breve{d}_u^{2}[n]}} e^{-j\pi \tau_u[n] f_{\text{c}}} e^{j2\pi f_u^{\mathrm{D}}[n]t_0} \nonumber \\
  & \times \bm{\alpha}_{\mathrm{tx}}\big( \pmb{c}_{\text{UAV}}[n],\breve{\pmb{c}}_{u}[n]\big)
\end{align}
where $G_{\mathrm{u}}^{\mathrm{rx}}$ denotes the receive antenna gain of buoy $u$ and $\breve{d}_u^{2}[n]=\left\|\pmb{c}_{\text{UAV}}[n]-\breve{\pmb{c}}_{u}[n]\right\|$.

\subsection{Buoy Uplink and UAV Backhaul Model}\label{S2.3}

After supplied with energy, buoys upload the monitoring data to the UAV, part of which are then relayed to the ship for timely data analysis. Let $x_u^{\text{c}}[n]$ and $x_u^{\text{b}}[n]$ be the data transmitted by buoy $u$ and the UAV at the $n$-th TS, respectively, with $\mathbb{E}\left\{\big|x^{\text{c}}_u[n]\big|^2\right\}\! =\!p_u[n]$ and $\mathbb{E}\left\{\big|x^{\text{b}}_u[n]\right|^2\big\}\! =\! 1$. Then the received signals at the UAV and ship can be expressed respectively in \eqref{eq:MDDSCC:ul} and \eqref{eq:MDDSCC:bk}:
\begin{align}\label{eq:MDDSCC:ul} 
  y_u^{\text{c}}[n] &= \tilde{\pmb{w}}_u^{\text{H}}[n]\breve{\bm{h}}^{\text{c}}_u[n] \alpha_u[n]\sqrt{p_u[n]}x_u^{\text{c}}[n] + \tilde{\pmb{w}}_u^{\text{H}}[n]\pmb{n}_{\text{c}}[n] , \\
  \label{eq:MDDSCC:bk}
  y_u^{\text{b}}[n] &= \left(\breve{\bm{h}}_u^{\text{b}}[n]\right)^{\text{H}} \tilde{\pmb{f}}_u[n]\alpha_u[n]x_u^{\text{b}}[n] + {n}_{\text{b}}[n] ,
\end{align}
where $\tilde{\pmb{w}}_u[n]$ and $\tilde{\pmb{f}}_u[n]$ are respectively the UAV's combiner and precoder vectors for receiving the uplink signal and transmitting the backhaul signal, and the corresponding channels, $\breve{\bm{h}}^{\text{c}}_u[n]$ and ${\bm{h}}^{\text{b}}_u[n]$, have the similar form with $\breve{\bm{h}}^{\text{p}}_u[n]$ as shown in \eqref{eq:MDDSCC:hpu}. The achieved data rates at the $n$-th TS over uplink and backhaul are given respectively by 
\begin{align} 
  R_u^{\text{c}}[n] =& \gamma_u^{\text{c}}[n] \log\bigg(1+\frac{\alpha_u[n]p_u[n]\big|\tilde{\pmb{w}}_u^{\text{H}}[n]\breve{\bm{h}}^{\text{c}}_u[n]\big|^2}{N_0\left\|\tilde{\pmb{w}}_u[n]\right\|^2}\bigg) , \label{eqULr} \\
  R_u^{\text{b}}[n] =& \gamma_u^{\text{b}}[n] \log\bigg(1+\frac{\alpha_u[n]\big|\big({\bm{h}}^{\text{b}}[n]\big)^{\text{H}} \tilde{\pmb{f}}[n]\big|^2}{N_0}\bigg). \label{eqBHr}
\end{align}

\section{Problem Formulation}\label{S3}

We deal with a practical scenario, in which the ship reaches a specific sea area and releases the UAV to collect the monitoring data from all the buoys therein. 
Since buoys used for monitoring in the open sea are energy-limited, the energy for implementing uplink transmissions is no larger than that is harvested during the power transferring stage. During the UAV mission period, the objective is to collect the monitoring data as much as possible. To avoid the data deficiency of any observation area, the amount of collected data from each buoy should be no less than a threshold. In light of the timeliness of monitoring data, the UAV promptly forwards part of the collected data to the central ship during flying. 

As the LoS path dominates the channel between any buoy and the UAV as well as only one buoy can be served by the UAV at each TS, the UAV adopts the matched-filter beamforming strategy to transmit/receive signals, i.e.,
\begin{align}\label{eq:MDDSCC:bf} 
  & \left\{ \begin{array}{l}
    \pmb{f}_u[n] = \pmb{v}_u[n] = \sqrt{P_{\text{UAV}}}\pmb{a}_{\text{tx}}\left(\pmb{c}_{\text{UAV}}[n],\pmb{c}_u[n]\right), \\
    \pmb{w}_u[n] = \tilde{\pmb{w}}_u[n] = \pmb{a}_{\text{rx}}\left(\pmb{c}_{\text{UAV}}[n],\pmb{c}_u[n]\right), \\ 
    \tilde{\pmb{f}}[n] = \sqrt{P_{\text{UAV}}}\pmb{a}_{\text{tx}}\left(\pmb{c}_{\text{UAV}}[n],\pmb{c}_{\text{ship}}\right) ,
  \end{array}\right.
\end{align} 
where $P_{\text{UAV}}$ is the UAV's maximum transmit power, and $\pmb{c}_{\text{ship}}$ is the 3D coordinate of the ship, which remains unchanged during mission period. The optimization problem to achieve the mission objective can be formulated as
\begin{subequations}\label{eq:MDDSCC:Opt} 
\begin{align}
  & (\text{P}1)\!:\!\!\!\!\!\! \max_{\left\{\alpha_{u}[n]\right\}, \left\{{\gamma}_u^{\text{x}}[n]\right\}, \left\{\bm{c}_{\text{UAV}}[n]\right\},\left\{p_u[n]\right\} }  \sum_{n=1}^{N}\sum_{u=1}^{U}R^{\text{c}}_u[n], \label{eqOPob} \\
   \text{s.t.} \, &\alpha_u[n]\in \left\{0,1\right\}, \sum_{u\in \mathcal{U}}\alpha_u[n]\leq 1, \forall n, \label{eqOPc1} \\
  &  0\leq\gamma_u^{\text{x}}[n]\leq 1, \sum_{\text{x}}\gamma_u^{\text{x}}[n]\leq 1, \forall n, \text{x} \in \{\text{s},\text{p},\text{c},\text{b}\}, \label{eqOPc2} \\
  & R_u^{\text{s}}[n]\geq \alpha_u[n]\Gamma_{\text{s}}^{\text{th}}, \forall u,n, \label{eqOPc3} \\
  & P_u^{\text{p}}[n]\geq \gamma_u^{\text{c}}[n]\alpha_u[n]p_u[n], \ \forall u,n, \label{eqOPc4} \\
  & \frac{1}{N}\sum_{n=1}^{N}R^{\text{c}}_u[n]\geq \Gamma_{\text{c}}^{\text{th}}, \forall u, \label{eqOPc5} \\
  & R^{\text{b}}_u[n]\geq \chi_u R^{\text{c}}_u[n], \forall u,n, \label{eqOPc6} \\ 
  & \left\|\bm{c}_{\text{UAV}}[n]-\bm{c}_{\text{UAV}}[n-1]\right\|\leq V_{\text{max}}T_{\text{t}}, \forall n>1 , \label{eqOPc7} \\
  & \bm{c}_{\text{UAV}}^{\text{I}}=\bm{c}_{\text{UAV}}[1], \ \bm{c}_{\text{UAV}}^{\text{F}}=\bm{c}_{\text{UAV}}[N] . \label{eqOPc9}
\end{align} 
\end{subequations}
Constraint \eqref{eqOPc1} specifies that only one buoy can be served by the UAV within each TS. Constraint \eqref{eqOPc2} states the time allocation among different stages, which may vary at different TSs. To meet the sensing QoS, \eqref{eqOPc3} states that the sensing MI of buoy $u$ within the $n$-th TS (if $\alpha_u[n]=1$) must be larger than a minimum estimation threshold $\Gamma_{\text{s}}^{\text{th}}$. Due to the limited service life of buoys, buoys only use the harvested power to carry out uplink transmission, as stated in \eqref{eqOPc4}. Constraint \eqref{eqOPc5} states that the collected data from each buoy must be larger than a rate threshold $\Gamma_{\text{c}}^{\text{th}}$. Constraint \eqref{eqOPc6} states that $\chi_u$ ($0\! <\! \chi_u\! <\! 1$) proportion of the collected data by buoy $u$ within each TS are offloaded to the ship. Constraints \eqref{eqOPc7} and \eqref{eqOPc9} specify the UAV's maximum flying distance within one TS, and its initial and final positions. 

\section{The Solutions to the Optimization Problem}\label{S4}

The proposed optimization (P1), jointly considering the user association between buoys and the UAV, time allocation among different tasks, buoys' transmit power and UAV's trajectory, is a mixed-integer non-convex problem. In particular, unlike the terrestrial UAV-assisted ISAC, the existence of sea clutters in the proposed maritime networks further introduces extreme difficulty to (P1).

To this end, we re-formulate the expressions of $R^{\text{s}}_u[n]$, $P^{\text{p}}_u[n]$, $R^{\text{c}}_u[n]$ and $R^{\text{b}}_u[n]$ using the beamforming strategy \eqref{eq:MDDSCC:bf}. Specifically, $R^{\text{s}}_u[n]$ is re-formulated as \eqref{eq:MDDSCC:Rus} at the bottom of previous page,
where $G_{\text{gain}}^{\text{s}}\! =\! P_{\text{UAV}}G_{\mathrm{UAV}}^{\mathrm{tx}} G_{\mathrm{UAV}}^{\mathrm{rx}}$ and $\bar{d}_u[n]\! =\! \sqrt{\left(x_{\text{UAV}}[n]\! -\! x_u[n]\right)^2\! +\! \left(y_{\text{UAV}}[n]\! -\! y_u[n]\right)^2}$ denotes the 2D Euclidean distance between the UAV and buoy $u$. Moreover, $R_u^{\text{c}}[n]$ is re-formulated as \setcounter{equation}{16}
\begin{align}\label{eq:MDDSCC:Ruc} 
  R_u^{\text{c}}[n] =& \gamma_u^{\text{c}}[n] \alpha_u[n] \log\left(1+\frac{p_u[n]G_{\text{gain}}^{\text{c}}\lambda^2}{N_0 4\pi^2 d_u^2[n]}\right),
\end{align}
the derivation of which is given in Appendix~\ref{App:MDDSCC:Ruc}, where $G_{\text{gain}}^{\text{c}}\! =\! G_u^{\mathrm{tx}} G_{\mathrm{UAV}}^{\mathrm{rx}}$. Similarly, $P_u^{\text{p}}[n]$ can be re-written as
\begin{align}\label{eq:MDDSCC:Pup} 
  P_u^{\text{p}}[n] =& \frac{\alpha_u[n]\gamma_u^{\text{p}}[n]\xi_u G_{\text{gain}}^{\text{p}}\lambda^2}{4\pi^2 d_u^2[n]} ,
\end{align} 
where $G_{\text{gain}}^{\text{p}}\! =\! P_{\text{UAV}}G_{\mathrm{UAV}}^{\mathrm{tx}} G_{\mathrm{u}}^{\mathrm{rx}}$. Also by defining $G_{\text{gain}}^{\text{b}}\! =\! P_{\text{UAV}}G_{\mathrm{UAV}}^{\mathrm{tx}} G_{\mathrm{ship}}^{\mathrm{rx}}$ and $d_{\text{b}}[n]\! =\! \left\|\pmb{c}_{\text{UAV}}[n]\! -\! \pmb{c}_{\text{ship}}\right\|$, $R_u^{\text{b}}[n]$ can be re-formulated as
\begin{align}\label{eq:MDDSCC:Rub} 
  R_u^{\text{b}}[n] =& \gamma_u^{\text{b}}[n] \alpha_u[n] \log\left(1+\frac{G_{\text{gain}}^{\text{b}}\lambda^2}{N_0 4\pi^2 d_{\text{b}}^2[n]}\right) .
\end{align}

\begin{remark}\label{Rm2}
As seen from \eqref{eq:MDDSCC:Ruc}-\eqref{eq:MDDSCC:Pup}, for moderate sea state and adequate UAV flight height, the effect of position deviation of buoys caused by sea waves on uplink communication and wireless charging within one TS is negligible. Even in severer sea state, the length of TS can be further shortened to guarantee that the fluctuation of buoys within one TS is limited in the range of UAV's half-power beamwidth. However, the position drift of buoys accumulates over TSs, leading to beam misalignment between the UAV and buoys. Hence the sensing stage at the beginning of each TS is essential for position recalibration.
\end{remark}

It can be found that the association indicators $\{\alpha_u[n]\}$ and the time allocation variables $\{\gamma_u^{\text{x}}[n]\}$ coexist inside expressions \eqref{eq:MDDSCC:Rus}-\eqref{eq:MDDSCC:Rub}, and they satisfy the  relationship:
\begin{equation}\label{eqAaG} 
\begin{cases}
  \gamma_u^{\text{s}}[n]+\gamma_u^{\text{p}}[n]+\gamma_u^{\text{c}}[n]+\gamma_u^{\text{b}}[n]=1,\  &\text{if} \  \alpha_u[n]=1, \\
  \gamma_u^{\text{s}}[n]+\gamma_u^{\text{p}}[n]+\gamma_u^{\text{c}}[n]+\gamma_u^{\text{b}}[n]=0,\  &\text{if} \ \alpha_u[n]=0 .
\end{cases}
\end{equation}
To deal with the tight coupling of $\{\alpha_u[n]\}$ and $\{\gamma_u^{\text{x}}[n]\}$, we divide (P1) into two sub-problems, which are given as
\begin{subequations}\label{eq:MDDSCC:Opt1.1} 
\begin{align}
  (\text{P}1.1): & \max_{\mathcal{S}_1 }  \sum_{n=1}^{N}\sum_{u=1}^{U}R^{\text{c}}_u[n], \label{Opt2:eqOPob} \\
  \text{s.t.}~~~ & \eqref{eqOPc1}, \eqref{eqOPc3}-\eqref{eqOPc9},
\end{align}
\end{subequations}
where $\mathcal{S}_1\! =\! \left\{\alpha_u[n],\bm{c}_{\text{UAV}}[n], p_u[n] \right\}$, which is also the set of optimization variables in (P2) of (\ref{eq:MDDSCC:Opt2}), and 
\begin{subequations}\label{eq:MDDSCC:Opt1.2} 
\begin{align}
  (\text{P}1.2): & \max_{\mathcal{S}_2} \sum_{n=1}^{N}\sum_{u=1}^{U}R^{\text{c}}_u[n], \label{Opt3:eqOPob} \\
  \text{s.t.}~~~ & \eqref{eqOPc2}-\eqref{eqOPc6},
\end{align}
\end{subequations}
where $\mathcal{S}_2\! =\! \left\{\gamma_u^{\text{x}}[n] \right\}$. 
Given $\mathcal{S}_2$, (P1.1) optimizes $\mathcal{S}_1$, the buoy-UAV association and UAV trajectory, while given $\mathcal{S}_1$, (P1.2) optimizes $\mathcal{S}_2$, the time allocation within each TS. Using alternating optimization, (P1.1) and (P1.2) are iteratively processed to finally achieve the optimal solution of (P1).
Obviously, once the set $\mathcal{S}_1$ is given, the convex optimization problem (P1.2) can be readily addressed. To solve (P1.1), however, its objective function as well as the involved constraints \eqref{eqOPc3}, \eqref{eqOPc4}, \eqref{eqOPc5} and \eqref{eqOPc6} must be transformed into tractable convex ones.

\subsection{Solution to (P1.1)}\label{S4.1}

\subsubsection{Optimization Strategy}

The binary variables $\left\{\alpha_u[n]\right\}$ hinder the convexification of the objective function and the related constraints in (P1.1). To deal with $\left\{\alpha_u[n]\right\}$, it is first converted into a continuous form, i.e., $0\! \le\! \alpha_u[n]\! \le\! 1$, and the slack continuous variables $\left\{\bar{\alpha}_u[n]\right\}$ are introduced to yielding two equivalent equality constraints given by
\begin{equation}\label{eq:MDDSCC:aun} 
\alpha_u[n](1-\bar{\alpha}_u[n])=0, \ \alpha_u[n]=\bar{\alpha}_u[n], \ \forall u,n .
\end{equation}
Hence, although $\alpha_u[n]$ is relaxed to a continuous form, it can only be 1 or 0 owing to the equality constraints \eqref{eq:MDDSCC:aun}.

After integrating the two equality constraints in \eqref{eq:MDDSCC:aun} into the objective function of (P1.1) as penalty terms, a new optimization problem (P2) is formed:
\begin{subequations}\label{eq:MDDSCC:Opt2} 
\begin{align}
  (\text{P}2)\!: & \! \max_{{\mathcal{S}}_1, \left\{\bar{\alpha}_u[n]\right\}}\! \sum_{n=1}^{N}\! \sum_{u=1}^{U}\! R^{\text{c}}_u[n]\! -\! \frac{1}{\eta}\! \sum_{n=1}^{N}\! \sum_{u=1}^{U}\! \Big(\! \left(\alpha_u[n]\left(1\! -\! \bar{\alpha}_u[n]\right)\right)^2 \nonumber \\
  & \hspace*{10mm}+\left(\alpha_u[n]-\bar{\alpha}_u[n]\right)^2\Big), \label{Opt2-1:eqOPob} \\
  \text{s.t.}~ & \eqref{eqOPc3}-\eqref{eqOPc9},
\end{align}
\end{subequations}
where $\eta\! >\! 0$ is a tunable parameter to control the optimization progress. As $\eta$ is gradually reduced to $0$, the optimization result of (P2) is equal to that of (P1.1). The optimization problem (P2) consists of several sets of variables, $\left\{\alpha_u[n]\right\}$, $\left\{\bar{\alpha}_u[n]\right\}$, $\left\{p_u[n]\right\}$ and $\left\{\bm{c}_{\text{UAV}}[n]\right\}$. Hence, it can be solved by iteratively fixing some of these sets and optimizing the remaining ones. 

As $\left\{\bar{\alpha}_u[n]\right\}$ only exist in the objective function and rely on the given $\left\{\alpha_u[n]\right\}$, we first derive the optimal $\left\{\bar{\alpha}_u[n]\right\}$ through differentiation as $\bar{\alpha}^{\text{opt}}_u[n]\! =\! \frac{\alpha_u[n] + \alpha_u^2[n]}{1 + \alpha_u^2[n]}$, $\forall u,n$. Given $\mathcal{S}_3\! =\! \left\{\alpha_u[n],\bar{\alpha}_u[n]\right\}$, the optimization of UAV trajectory and buoys' transmit power is formulated as
\begin{subequations}\label{eq:MDDSCC:Opt2.1} 
\begin{align}
  (\text{P}2.1): & \max_{{\mathcal{S}}_4} \sum_{n=1}^{N}\sum_{u=1}^{U}R^{\text{c}}_u[n], \label{Opt2.1:eqOPob} \\
  \text{s.t.}~~~ & \eqref{eqOPc3}-\eqref{eqOPc9},
\end{align}
\end{subequations} 
where ${\mathcal{S}}_4\! =\! \left\{\bm{c}_{\text{UAV}}[n], p_u[n] \right\}$. Given ${\mathcal{S}}_4$ and $\left\{\bar{\alpha}_u[n]\right\}$, the optimization of $\left\{\alpha_u[n]\right\}$ can be readily solved, as the objective function is concave and the involved constraints are linear. Obviously, the challenge of solving the optimization problem (P2) resides in dealing with (P2.1), since the objective function and most of constraints of (P2.1) are non-convex. Therefore, we address the convexification of the objective function and constraints of (P2.1).

\subsubsection{Objective Function Convexification for Sub-Problem (P2.1)}

The objective function of (P2.1) has a form of $\sum_{n=1}^N\! \sum_{u=1}^{U}\! \alpha_u[n]\log\big(1\! +\! \frac{\mathcal{A}_u[n]}{\mathcal{B}_u[n]}\big)$, where $\mathcal{A}_u[n]\! =\! p_u[n]G_{\text{gain}}^{\text{c}}\lambda^2$ and $\mathcal{B}_u[n]\! =\! N_0 4\pi^2 d_u^2[n]$ are both linear and convex functions. According to the principle of quadratic transform \cite{shen2018fractional}, it is equivalent to 
$\sum_{n=1}^N\! \sum_{u=1}^{U}\! \alpha_u[n]\log(1\! +\! 2\varpi_u[n]\sqrt{\mathcal{A}_u[n]}\! -\! \varpi_u^2[n]\mathcal{B}_u[n])$, which is a concave function, when either ${\mathcal{S}}_4$ or $\varpi_u[n]\in \mathbb{R}$ is fixed.

\begin{figure*}[!t]\setcounter{equation}{27}
\vspace*{-1mm}
\begin{align}\label{eq:MDDSCC:LamdaC} 
  \Lambda_u\left(\bm{c}_{\text{UAV}}[n]\right) \approx& \tilde{\Lambda}_u\left(\bm{c}_{\text{UAV}}[n]\right) \triangleq \Lambda_u\left(\bm{c}_{\text{UAV}}^{(k-1)}[n]\right) + {\frac{\partial \Lambda _u\left( \boldsymbol{c}_{\mathrm{UAV}}\left[ n \right] \right)}{\partial x_{\mathrm{UAV}}\left[ n \right]}}\bigg|_{\boldsymbol{c}_{\mathrm{UAV}}^{\left( k-1 \right)}\left[ n \right]}\left( x_{\mathrm{UAV}}\left[ n \right] -x_{\mathrm{UAV}}^{\left( k-1 \right)}\left[ n \right] \right)\nonumber \\ 
 & +{\frac{\partial \Lambda _u\left( \boldsymbol{c}_{\mathrm{UAV}}\left[ n \right] \right)}{\partial y_{\mathrm{UAV}}\left[ n \right]}}\bigg|_{\boldsymbol{c}_{\mathrm{UAV}}^{\left( k-1\right)}\left[ n \right]}\left( y_{\mathrm{UAV}}\left[ n \right] -y_{\mathrm{UAV}}^{\left( k-1 \right)}\left[ n \right] \right) ,
\end{align}
\hrulefill
\begin{align}\label{eq:MDDSCC:PiC} 
  & \Pi_u\left(\bm{c}_{\text{UAV}}[n],p_u[n]\right)\approx \tilde{\Pi}_u\left(\bm{c}_{\text{UAV}}[n],p_u[n]\right)\triangleq \Pi_u\left(\bm{c}^{(k-1)}_{\text{UAV}}[n],p^{(k-1)}_u[n]\right)+ {\frac{\partial \Pi_u\left(\bm{c}_{\text{UAV}}[n],p_u[n]\right)}{\partial x_{\mathrm{UAV}}\left[ n \right]}}\bigg|_{\boldsymbol{c}_{\mathrm{UAV}}^{\left( k-1 \right)}\left[ n \right],p^{(k-1)}_u[n]} \nonumber\\
  &\hspace*{20mm} \times \left( x_{\mathrm{UAV}}\left[ n \right] -x_{\mathrm{UAV}}^{\left( k-1 \right)}\left[ n \right] \right)
+{\frac{\partial \Pi_u\left(\bm{c}_{\text{UAV}}[n],p_u[n]\right)}{\partial y_{\mathrm{UAV}}\left[ n \right]}}\bigg|_{\boldsymbol{c}_{\mathrm{UAV}}^{\left( k-1 \right)}\left[ n \right],p^{(k-1)}_u[n]} \left( y_{\mathrm{UAV}}\left[ n \right] -y_{\mathrm{UAV}}^{\left( k-1 \right)}\left[ n \right] \right) \nonumber \\
  &\hspace*{20mm} + {\frac{\partial \Pi_u\left(\bm{c}_{\text{UAV}}[n],p_u[n]\right)}{\partial p_u[n]}}\bigg|_{\boldsymbol{c}_{\mathrm{UAV}}^{\left( k-1 \right)}\left[ n \right],p^{(k-1)}_u[n]}\left(p_u[n]-p^{(k-1)}_u[n]\right) ,
\end{align}
\hrulefill
\begin{align}\label{eq:MDDSCC:LamdaCx} 
  & \left\{ \begin{array}{l}
    \frac{\partial \Lambda _u\left(\bm{c}_{\mathrm{UAV}}\left[ n \right] \right)}{\partial x_{\mathrm{UAV}}\left[ n \right]} = \left(C_1 e^{-\frac{\Gamma_s \bar{d}^2_u[n]}{\bar{z}}}\left(-\frac{2 \Gamma_s\left(\bar{d}^2_u[n] + \bar{z}\right)}{\bar{z}} + \frac{\bar{d}^2_u[n] - \bar{z}}{\bar{d}^2_u[n]}\right) + 4 C_2 \bar{d}_u[n]\left(\bar{d}^2_u[n] + \bar{z}\right)\right)\left(\frac{x_{\mathrm{UAV}}[n] - x_u[n]}{\bar{d}_u[n]}\right) , \\
    \frac{\partial \Pi_u\left(\bm{c}_{\text{UAV}}[n],p_u[n]\right)}{\partial x_{\mathrm{UAV}}\left[ n \right]} = {-2 r^{\text{bc}}_u[n]} \left( 1 + \frac{C_3p_u[n]}{d_u^2[n]}\right)^{r^{\text{bc}}_u[n]-1} \frac{C_3p_u[n]\left(x_{\mathrm{UAV}}[n] - x_u[n]\right)}{d_u^4[n]} - \frac{2C_4}{d_{\text{b}}^4[n]}\left(x_{\mathrm{UAV}}[n] - x_{\text{b}}[n]\right) , \\
    \frac{\partial \Pi_u\left(\bm{c}_{\text{UAV}}[n],p_u[n]\right)}{\partial p_u\left[ n \right]} = {r^{\text{bc}}_u[n]}\left(1+\frac{C_3p_u[n]}{d_u^2[n]}\right)^{r^{\text{bc}}_u[n]-1} \frac{C_3}{d_u^2[n]} .
  \end{array}\right. 
\end{align}
\hrulefill
\vspace*{-4mm}
\end{figure*}

\begin{figure*}[!b]\setcounter{equation}{34}
\vspace*{-4mm}
\hrulefill
\begin{align}\label{eq:MDDSCC:Rucc} 
  R_u^{\text{c}}[n] \approx & \tilde{R}_u^{\text{c}}[n] \triangleq R_u^{\text{c}(k-1)}[n] + {\frac{\partial R_u^{\text{c}}[n]}{\partial x_{\mathrm{UAV}}\left[ n \right]}}\bigg|_{\boldsymbol{c}_{\mathrm{UAV}}^{\left( k-1 \right)}[ n ],p^{(k-1)}_u[n]}\left( x_{\mathrm{UAV}}\left[ n \right] -x_{\mathrm{UAV}}^{\left( k-1 \right)}\left[ n \right] \right) \nonumber \\ 
  & + {\frac{\partial R_u^{\text{c}}[n]}{\partial y_{\mathrm{UAV}}\left[ n \right]}}\bigg|_{\boldsymbol{c}_{\mathrm{UAV}}^{\left( k-1\right)}\left[ n \right],p^{(k-1)}_u[n]}\left( y_{\mathrm{UAV}}\left[ n \right] -y_{\mathrm{UAV}}^{\left( k-1 \right)}\left[ n \right] \right) + {\frac{\partial R_u^{\text{c}}[n]}{\partial p_u[n]}}\bigg|_{\boldsymbol{c}_{\mathrm{UAV}}^{\left( k-1 \right)}\left[ n \right],p^{(k-1)}_u[n]}\left(p_u[n]-p^{(k-1)}_u[n]\right) ,
\end{align}
\vspace*{-2mm}
\end{figure*} 

\subsubsection{Constraints Convexification for Sub-Problem (P2.1)}

To deal with the non-convex constraints \eqref{eqOPc3} and \eqref{eqOPc6} with $\alpha_u[n]\! \neq\! 0$, we first rewrite them respectively as\setcounter{equation}{25} 
\begin{align} 
  & {C}_{1} e^{-\frac{\Gamma_{\text{s}}{\bar{d}^2_u[n]}}{\bar{z}}}\! \left(\frac{\bar{d}^2_u[n]\! +\! \bar{z}}{\bar{d}_u[n]}\right)\! +\! C_{2}\left(\bar{d}_u^2[n]\! +\! \bar{z}\right)^2\! \leq C_{u}[n],\! \label{eq:MDDSCC:T15d} \\
  & \left(1+\frac{C_3p_u[n]}{d_u^2[n]}\right)^{r^{\text{bc}}_u[n]} - \left(1+\frac{C_4}{d^2_{\text{b}}[n]}\right) \leq 0 , \label{eq:MDDSCC:T15g}
\end{align}
where $\bar{z}\! =\! {\left(z_{\text{UAV}}\! -\! z_u[n]\right)^2}$, $C_{u}[n]\! =\! \frac{G_{\text{gain}}^{\text{s}}\lambda^2 \psi_u R B}{2^{\left({\Gamma_{\text{s}}^{\text{th}}}/{\gamma_u^{\text{s}}[n]}\right)}-1}$, $C_{1}\! =\! 0.886c\, G_{\text{gain}}^{\text{s}} \lambda^2 \Gamma_{\text{s}}$, $C_{2}\! =\! N_0 R B 4\pi^3$, $C_3\! =\! \frac{G_{\text{gain}}^{\text{c}}\lambda^2}{N_0 4\pi^2}$, $C_4\! =\! \frac{G_{\text{gain}}^{\text{b}}\lambda^2}{N_0 4\pi^2}$ and $r^{\text{bc}}_u[n]\! =\! \frac{\gamma_u^{\text{c}}[n]\chi_u}{\gamma_u^{\text{b}}[n]}$. 
Define the left-hand-sides of \eqref{eq:MDDSCC:T15d} and \eqref{eq:MDDSCC:T15g} as $\Lambda_u\left(\bm{c}_{\text{UAV}}[n]\right)$ and $\Pi_u\left(\bm{c}_{\text{UAV}}[n],p_u[n]\right)$, respectively, 
which can be iteratively approximated using the first-order Taylor expansion (FOTE) at local point $\bm{c}^{(k)}_{\text{UAV}}[n]$ and $p^{(k)}_u[n]$, and hence they can be rewritten as \eqref{eq:MDDSCC:LamdaC} and \eqref{eq:MDDSCC:PiC} at the top of this page, where $\frac{\partial \Lambda _u\left( \boldsymbol{c}_{\mathrm{UAV}}\left[ n \right] \right)}{\partial x_{\mathrm{UAV}}\left[ n \right]},\frac{\partial \Pi_u\left(\bm{c}_{\text{UAV}}[n],p_u[n]\right)}{\partial x_{\mathrm{UAV}}\left[ n \right]}$ and $\frac{\partial \Pi_u\left(\bm{c}_{\text{UAV}}[n],p_u[n]\right)}{\partial p_u[n]}$ are given in \eqref{eq:MDDSCC:LamdaCx}. 
In addition, the similar expressions of $\frac{\partial \Lambda _u\left( \boldsymbol{c}_{\mathrm{UAV}}\left[ n \right] \right)}{\partial y_{\mathrm{UAV}}\left[ n \right]}$ and $\frac{\partial \Pi_u\left(\bm{c}_{\text{UAV}}[n],p_u[n]\right)}{\partial y_{\mathrm{UAV}}\left[ n \right]}$ are omitted for the sake of conciseness. 
As long as the values of $\big|\bm{c}_{\text{UAV}}[n]-\bm{c}_{\text{UAV}}^{(k-1)}[n]\big|$ and $\big|p_u[n]-p_u^{(k-1)}[n]\big|$ are sufficiently small, the constraints \eqref{eqOPc3} and \eqref{eqOPc6} can be substituted respectively by \setcounter{equation}{30}
\begin{align} 
  & \tilde{\Lambda}_u\left(\bm{c}_{\text{UAV}}[n]\right)-C_u[n] \leq 0, \forall u,n, \label{eq:MDDSCC:SubeqOPc3} \\
  & \tilde{\Pi}_u\left(\bm{c}_{\text{UAV}}[n],p_u[n]\right) \leq 0 , \forall u,n. \label{eq:MDDSCC:SubeqOPc6}
\end{align}

Next, when $\alpha_u[n]\! \neq\! 0$, the constraint \eqref{eqOPc4} can be re-expressed as $d_u^2[n]\! -\! \frac{J_u[n]}{p_u[n]}\! \leq\! 0$, which has a form of convex-minus-convex, and $J_u[n]\! =\! \frac{\gamma_u^{\text{p}}[n]\xi_u G_{\text{gain}}^{\text{p}}\lambda^2}{\gamma_u^{\text{c}}[n] 4\pi^2}$. We can resort to the inner convex approximation method and obtain the concave lower bound of $\frac{J_u[n]}{p_u[n]}$, which is given by \cite{marks1978general}
\begin{align}\label{eqICAM} 
  \frac{J_u[n]}{p_u[n]} \geq&  \frac{2J_u[n]}{p_u^{(k)}[n]}-\frac{p_u[n]J_u[n]}{\big(p_u^{(k)}[n]\big)^2} .
\end{align}
Consequently, the constraint \eqref{eqOPc4} can be iteratively replaced with the following convex one
\begin{align}\label{eq:MDDSCC:SubeqOPc4} 
d_u^2[n]-\frac{2J_u[n]}{p_u^{(k)}[n]}+\frac{p_u[n]J_u[n]}{(p_u^{(k)}[n])^2} \leq 0, \ \forall u,n .
\end{align}

Although the constraint \eqref{eqOPc5} is mainly composed of $R_u^{\text{c}}[n]$, the quadratic transform used to convexify the objective function of (P2.1) is inapplicable to it, since the initial iteration value may not satisfy the constraint. With the aid of FOTE, $R_u^{\text{c}}[n]$ is approximately transformed into \eqref{eq:MDDSCC:Rucc} at the bottom of this page, where \setcounter{equation}{35}
\begin{align}\label{eqR_uFOTE} 
  &\hspace*{-4mm}\left\{\!\! \begin{array}{l}
	  {\frac{\partial R_u^{\text{c}}[n]}{\partial x_{\mathrm{UAV}}\left[ n \right]}}=\frac{-2\gamma_u^{\text{c}}[n]\alpha_u[n]p_u[n]G_{\text{gain}}^{\text{c}}\lambda^2\left( x_{\mathrm{UAV}}\left[ n \right] -x_u\left[ n \right] \right)}{\ln^2(N_0 4\pi^2 d_u^4[n]+p_u[n]G_{\text{gain}}^{\text{c}}\lambda^2d_u^2[n])} , \\
    {\frac{\partial R_u^{\text{c}}[n]}{\partial p_u[n]}}=\frac{\gamma_u^{\text{c}}[n]\alpha_u[n]G_{\text{gain}}^{\text{c}}\lambda^2}{\ln^2(N_0 4\pi^2 d_u^2[n]+p_u[n]G_{\text{gain}}^{\text{c}}\lambda^2)} ,
  \end{array}\right.
\end{align}
while the similar form of $ \frac{\partial R_u^{\text{c}}[n]}{\partial y_{\mathrm{UAV}}\left[ n \right]}$ is omitted. Then, \eqref{eqOPc5} can be substituted with
\begin{align}\label{eq:MDDSCC:SubeqOPc5} 
\Gamma_{\text{c}}^{\text{th}} - \frac{1}{N}\sum_{n=1}^{N}\tilde{R}_u^{\text{c}}[n] \leq 0 , \forall u.
\end{align}
Up to this point, all the non-convex constraints have been transformed into the convex ones, and we can proceed to carry out the optimization implementation.

\subsubsection{Implementation of Sub-Problem (P2.1)} 

After the transformation of objective function and constraints, the optimization problem (P2.1) is equivalent to
\begin{subequations}\label{eq:MDDSCC:Opt2.2} 
\begin{align}
  ({\text{P}2.2}): & \max_{{\mathcal{S}}_4, \left\{\varpi_u[n]\right\}}  
\sum_{n=1}^N \sum_{u=1}^{U} \alpha_u[n]\log\big(1 + 2\varpi_u[n]\sqrt{\mathcal{A}_u[n]} \nonumber \\
  & \hspace*{28mm} - \varpi_u^2[n]\mathcal{B}_u[n]\big), \label{Opt2.2:eqOPob} \\
  \text{s.t.}~~~ & \eqref{eq:MDDSCC:SubeqOPc3}, \eqref{eq:MDDSCC:SubeqOPc6}, \eqref{eq:MDDSCC:SubeqOPc4}, \eqref{eq:MDDSCC:SubeqOPc5}, \eqref{eqOPc7}, \eqref{eqOPc9}, \\
  & \left|\bm{c}_{\text{UAV}}[n]-\bm{c}_{\text{UAV}}^{(k-1)}[n]\right| \leq \vartheta^{(k)}\upsilon_{\text{c}}, \ \forall n, \label{Opt2.2:eqOPoc} \\
  &\left|p_u[n]-p_u^{(k-1)}[n]\right| \leq \vartheta^{(k)}\upsilon_{\text{p}}, \ \forall u, n, \label{Opt2.2:eqOPod}
\end{align} 
\end{subequations} 
where $\vartheta^{(k)}$ is the adjustable factor to control the optimization step size at the $k$-th iteration, $\upsilon_{\text{c}}$ and $\upsilon_{\text{p}}$ are the unique scales of $\bm{c}_{\text{UAV}}[n]$ and $p_u[n]$, respectively. Since either ${\mathcal{S}}_4$ or $\left\{\varpi_u[n]\right\}$ is fixed, the objective function is concave, and given $\mathcal{S}_4$, the optimal $\varpi_u^{(k)}[n]$ is $\varpi_u^{(k)}[n]\! =\! \frac{\sqrt{\mathcal{A}_u^{(k-1)}[n]}}{\mathcal{B}_u^{(k-1)}[n]}, \forall n,u$. Therefore, the objective function can be iteratively updated during the optimization process.

However, to start the optimization procedure, a feasible initial point $\left(\bm{c}_{\text{UAV}}^{(0)}[n],p_u^{(0)}[n]\right)$ is required. In particular, it is impractical to randomly initialize $p_u^{(0)}[n]$, as each buoy's transmit power, i.e., $p_u^{(0)}[n]$, is harvested during UAV wireless charging, and the amount of which heavily depends on the UAV trajectory. Hence, the initialization should be done in the way that satisfies the constraints in (\ref{eq:MDDSCC:Opt2.2}b). Thus, a feasible initial point can be obtained by solving the following optimization
\begin{subequations}\label{eq:MDDSCC:Optini} 
\begin{align}
  (\underline{\text{P}2.2}): & \min_{\pmb{\omega}} \sum_{u=1}^U \bigg( \sum_{n=1}^N \big(\omega_{1,u}[n] + \omega_{2,u}[n] + \omega_{3,u}[n]\big) \nonumber \\
	& \hspace*{15mm} + \omega_{4,u} \bigg) , \label{Opt2.22:eqOPob} \\
  \text{s.t.}~~~~ & \eqref{eqOPc7}, \eqref{eqOPc9}, \\
  & \tilde{\Lambda}_u\left(\bm{c}_{\text{UAV}}[n]\right)-C_u[n] \leq\omega_{1,u}[n], \, \forall u, n, \label{eq39c} \\
  & \tilde{\Pi}_u\left(\bm{c}_{\text{UAV}}[n],p_u[n]\right) \leq \omega_{2,u}[n], \forall u,n, \label{eq39d} \\
  & d_u^2[n]-\frac{2J_u[n]}{p_u^{(k-1)}[n]}+\frac{p_u[n]J_u[n]}{(p_u^{(k-1)}[n])^2} \leq \omega_{3,u}[n], \forall u,n, \label{eq39de} \\
  & \Gamma_{\text{c}}^{\text{th}} - \frac{1}{N}\sum_{n=1}^{N}\tilde{R}_u^{\text{c}}[n] \leq \omega_{4,u}, \, \forall u, \label{eq39f} \\
  & \omega_{i,u}[n]\geq 0, \, 1\le i\le 3, \,\omega_{4,u}\geq 0, \,\forall u,n , \label{eq39g}
\end{align}
\end{subequations} 
where $\pmb{\omega}\! =\! \left\{\{\omega_{1,u}[n],\omega_{2,u}[n],\omega_{3,u}[n]\},\omega_{4,u}\right\}$. It can be seen that the feasible initial point $\left(\bm{c}_{\text{UAV}}^{(0)}[n],p_u^{(0)}[n]\right)$ is found when the objective value is close to zero.

\subsection{Algorithmic Analysis for Solving (P1)}\label{S4.2}

The optimization process of solving (P1) is presented in Algorithm~\ref{MDDSCC:al1}, which is based on alternating optimization and is composed of outer- and inner-layer iterations. Let $i$ and $j$ be the indexes of the outer- and inner-layer iterations, respectively, and denote $\mathcal{R}\! =\! \sum_{n=1}^{N}\sum_{u=1}^{U}R^{\text{c}}_u[n]$. Within each inner-layer iteration, the alternating update among $\left\{\alpha_u[n]\right\},\left\{\bar{\alpha}_u[n]\right\}$ and $\left\{p_u[n], \bm{c}_{\text{UAV}}[n]\right\}$ leads to an increased rate, i.e., $\mathcal{R}^{(j_1)}\! \leq\! \mathcal{R}^{(j_2)}$, $j_1\! \leq\! j_2$. After the inner-layer iteration, the update of time allocation further enhances the rate performance, leading to $\mathcal{R}^{(i_1)}\! \leq\! \mathcal{R}^{(j_1)}\! \leq\! \dots\! \leq\! \mathcal{R}^{(j_{\text{end}})}\! \leq\! \mathcal{R}^{(i_2)}$, $i_1\! \leq\! i_2$. The convergence behavior of the adopted quadratic transform and inner convex approximation have been demonstrated in \cite{polyak1998convexity,marks1978general}, and hence the convergence of Algorithm~\ref{MDDSCC:al1} is guaranteed.

\begin{algorithm}[t!]
\caption{Iterative Solution to Optimization (P1) } 
\label{MDDSCC:al1} 
\textbf{Initialization:} \\
  Give buoys' positions $\pmb{c}_u[n], \forall u, n$\;
	Initialize $\gamma_u[n], \alpha_u[n], \forall u, n$\;
  \DT{}{
	\Repeat{\em{Objective value of (P1) converges}}{
	For current $\mathcal{S}_2$, obtain optimal $\mathcal{S}_1$ by solving (P1.1) within following inner layer iteration\;
	\LT{}{
    \Repeat{\em{Objective value of (P2) converges}}{
    Compute $\bar{\alpha}_u[n]=\frac{\alpha_u[n]+\alpha_u^2[n]}{1+\alpha_u^2[n]},\ \forall u,n$\;
		Find initial feasible iteration values $\bm{c}_{\text{UAV}}^{(0)}[n]$ and $p_u^{(0)}[n]$ by solving $\underline{(P2.2)}$, and obtain optimal $\bm{c}_{\text{UAV}}[n]$ and $p_u[n]$ through (P2.2)\;
		Derive optimal $\alpha_u[n]$ when fixing $\mathcal{S}_4$ and $\bar{\alpha}_u[n]$ in (P2)\;
    }
  }
	For current $\mathcal{S}_1$, obtain optimal $\mathcal{S}_2$ by solving (P1.2);
  }
	}
  \KwOut{$\left\{\gamma_u^{*}[n],\alpha_u^{*}[n],\bm{c}_{\text{UAV}}^{*}[n], p_u^{*}[n]\right\}$.}
\end{algorithm}

\begin{remark}
To deal with (P2.2), the FOTE can be adopted to convexify the non-convex constraints. However, when the trust region, i.e., $\big|\bm{c}_{\text{UAV}}[n]\! -\! \bm{c}_{\text{UAV}}^{(k-1)}[n]\big|$ and $\big|p_u[n]\! -\! p_u^{(k-1)}[n]\big|$, is large, a problem may arise that even though the transformed constraints, e.g., \eqref{eq:MDDSCC:SubeqOPc3}, \eqref{eq:MDDSCC:SubeqOPc6}, \eqref{eq:MDDSCC:SubeqOPc4} and \eqref{eq:MDDSCC:SubeqOPc5}, are satisfied and (P2.2) is successively solved, the corresponding original constraints, e.g., \eqref{eqOPc3}-\eqref{eqOPc6}, are violated. We introduce an augmented FOTE (A-FOTE) to address this issue. Specifically, the adjustable factor $\vartheta^{(k)}$ is initialized with a sufficiently big value to enlarge the trust region at the beginning of the iteration. Following each execution of (P2.2), $\vartheta^{(k)}$ is adaptively decreased for the subsequent iteration if either: (i) no feasible solution exists, or (ii) a feasible solution is found but violates any of the original constraints. Using this A-FOTE, the inner layer iteration is guaranteed to converge in a trusted way.
\end{remark}

According to \cite{nguyen2019joint}, the computational complexity of each inner-layer iteration is given by $L_1\! =\! \mathcal{O}\big((6 N\! +\! U)^{2.5}(2 N)^2\! +\! (3 N\! +\! U)^4\big)$, and the computational complexity of optimizing (P1.2) is given by $L_2\! =\! \mathcal{O}\big((4 N\! +\! U)^{2.5}(4 N)^2\big)$. Hence, the total computational complexity of Algorithm~\ref{MDDSCC:al1} is $\rho_{\text{o}}\big(\rho_{\text{i}}L_1\! +\! L_2\big)$, where $\rho_{\text{i}}$ and $\rho_{\text{o}}$ denote the numbers of inner- and outer-layer iterations, respectively. 

\section{Simulation Results}\label{S5}

We evaluate the performance of the UAV-enabled ISCPB in maritime monitoring networks, and demonstrate the effectiveness of the proposed joint optimization method for UAV trajectory, time allocation, user association and power scheduling. In the simulation, we consider a practical buoy-based monitoring scenario in open sea, where $10$ buoys are uniformly distributed along a circular arc with a radius of 500\,m and centered at the coordinate of (500, 0, 0)\,m to form a buoy observing array. The ship that establishes the backhaul link with the UAV situates at (500, -20, 0)\,m. The UAV equipped with an $8 \times 8$ UPA antenna array takes off from the coordinate of (0, -10, 10)\,m and flies toward the destination point of (1000, -10, 10)\,m with a maximum speed of 30\,m/s. The whole mission period is discretized into 100 TSs, each of which contributes one second. Unless otherwise specified, the default system's parameters listed in Table~\ref{Table:MDDSCC:para} are used.

\begin{table}[t]
\caption{Default simulation system parameters}
\vspace*{-1mm}
\scriptsize
\centering
\begin{tabular}{l|l}
\hline
Parameters & Values \\ \hline
UAV power budget ($P_{\text{UAV}}$) & $36$ dBm \\ \hline
Antenna gains of ship, UAV and buoys & (30, 26, 20) dBi \\ \hline 
Noise power spectrum density ($N_0$) & -107 dBm \\ \hline
Energy conversion efficiency ($\xi_u, \ \forall u$) & 0.8 \\ \hline
Sea State ($\kappa_{\text{s}}$) & 3 \\ \hline
UAV flight height ($z_{\text{UAV}}$) & 10 m \\ \hline
Central frequency and bandwidth & $5$ GHz, 10 MHz \\ \hline
Minimum communication QoS ($\Gamma_{\text{c}}^{\text{th}}$) & 1 bps/Hz \\ \hline
Ratio of data offloaded to the ship ($\chi_u, \ \forall u$) & 0.2 \\ \hline
Minimum sensing QoS ($\Gamma_{\text{s}}^{\text{th}}$) & 1 bps/Hz \\ \hline
\end{tabular}
\label{Table:MDDSCC:para} 
\vspace*{-3mm}
\end{table}

\begin{figure}[!h]
\vspace*{-5mm}
\centering
\includegraphics[width=0.8\linewidth]{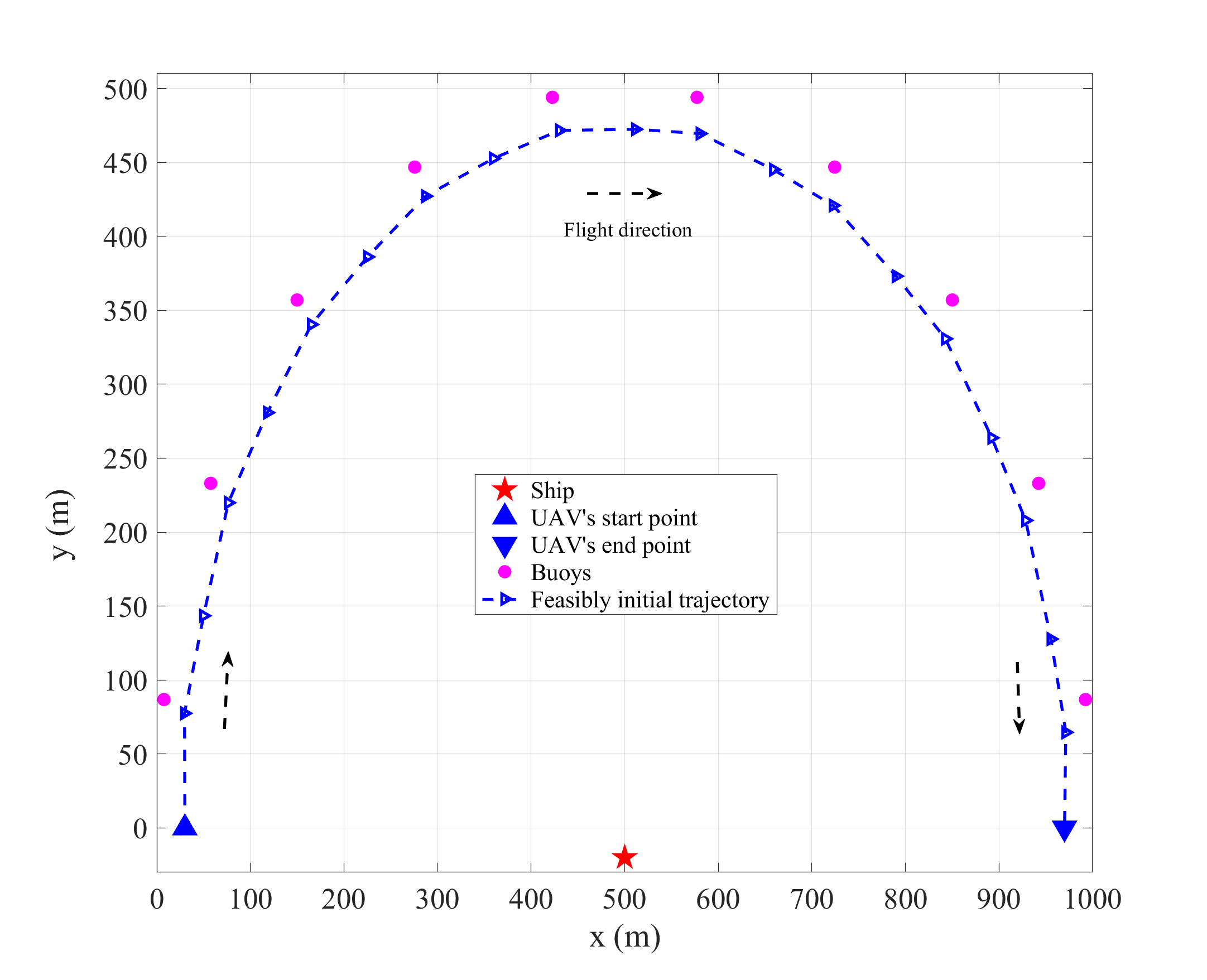}
\vspace*{-3mm}
\caption{\small The diagram of the feasibly initial UAV trajectory.}
\label{figure-MDDSCC-Simscen} 
\vspace*{-5mm}
\end{figure}

\subsection{Performance of Proposed Optimization Method}\label{S5.1}

To demonstrate the effectiveness of our proposed method for jointly optimizing the UAV trajectory, time allocation, user association and power scheduling, we evaluate its performance in terms of the achievable rate and sensing MI as well as the convergence behavior. First, we consider the optimization problem ($\underline{\text{P}2.2}$) and obtain a feasibly initial UAV trajectory, as shown in Fig.~\ref{figure-MDDSCC-Simscen}, that meets the basic QoSs of communication rates for both uplink and backhaul link as well as sensing. It can be observed that as the UAV is only connected with one buoy within each TS, its initial flight trajectory is generally along the circular arc following the buoys' coordinate distribution, such that the link between the UAV and the buoy being served is as short as possible. 

\begin{figure}[t]
\vspace*{-6mm}
\centering
\includegraphics[width=0.8\linewidth]{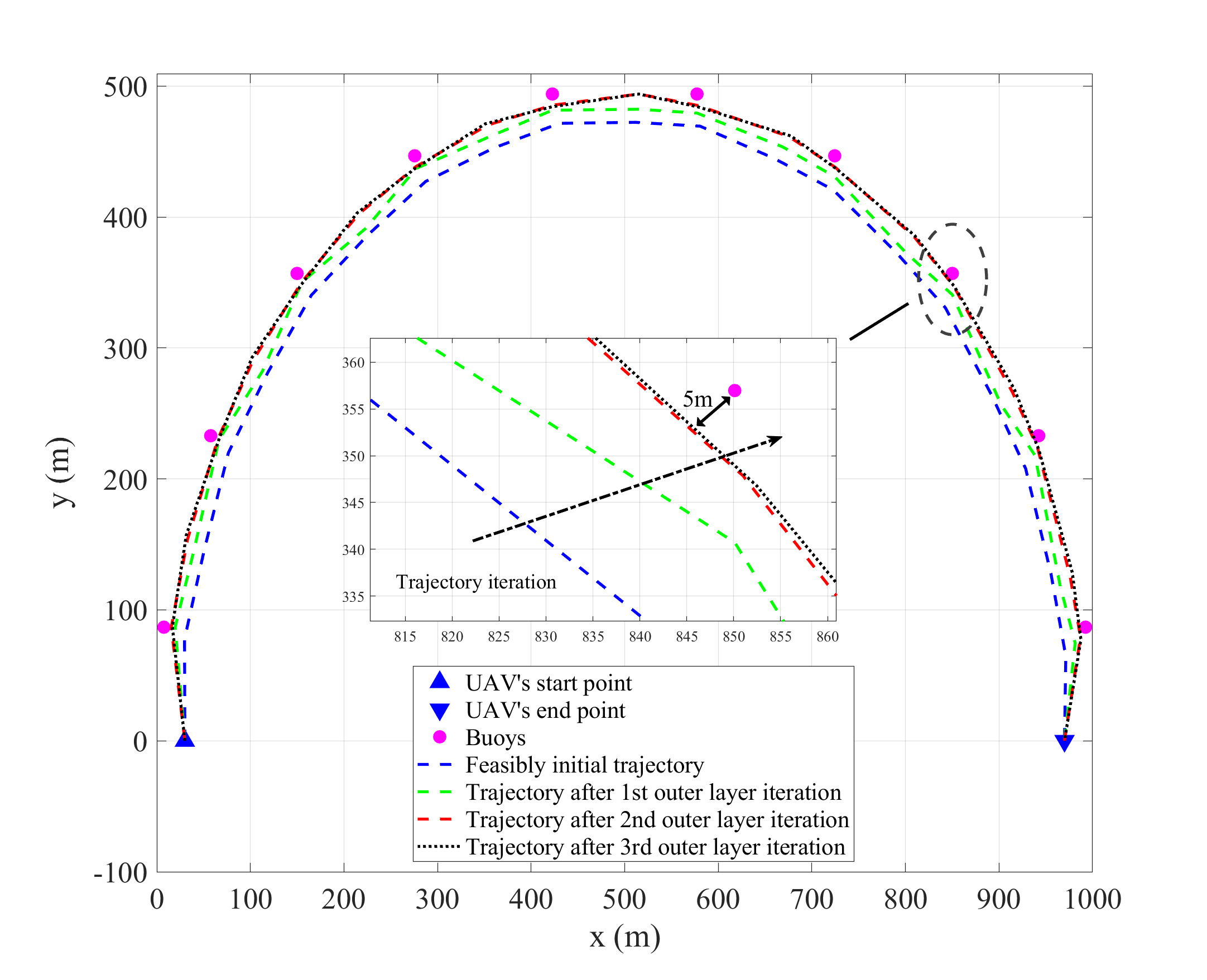}
\vspace*{-4mm}
\caption{\small The outer layer iteration process of UAV trajectory.}
\label{figure-MDDSCC-TracIter}  
\vspace*{-3mm}
\end{figure}

Then we examine the optimization process presented in Algorithm~\ref{MDDSCC:al1}. Specifically, the outer-layer iteration process of UAV trajectory is depicted in Fig.~\ref{figure-MDDSCC-TracIter}, which shows that three outer-layer iterations are needed to achieve convergence, and as the number of iterations increases, the UAV flies closer to buoys. However, although the shorter distance usually leads to better performance of sensing and communication, the UAV avoids to fly over the buoys. This is because the larger the grazing angle is, the stronger the sea clutter interference becomes, and to decrease the effect of sea clutter, the UAV keeps a feasible horizontal angle with buoys. Also it can be seen from Fig.~\ref{figure-MDDSCC-TracIter} that after the third iteration, the shortest distance between the UAV and buoy is slightly over 5\,m leading to around $60^{\circ}$ grazing angle. This exactly matches the results in Fig.~\ref{figure-MDDSCC-sc} which show that when $\kappa_{\text{s}}\! =\! 3$, once the grazing angle is larger than $60^{\circ}$, the effect of sea clutter becomes more severe.

\begin{figure}[!h]
\vspace*{-4mm}
\centering
\includegraphics[width=0.9\linewidth]{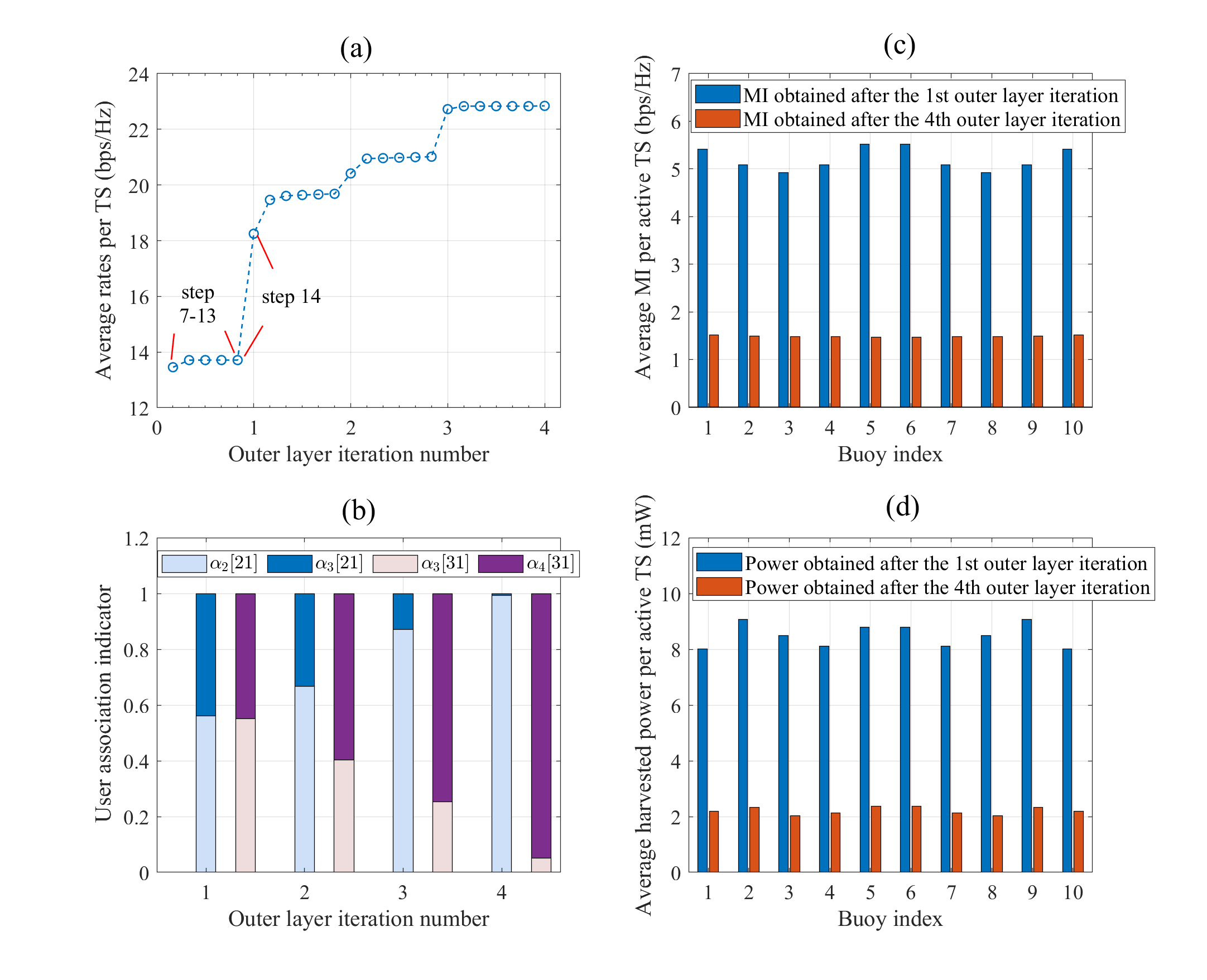}
\vspace*{-5mm}
\caption{\small The optimization performance in terms of (a) communication rate, (b) user association, (c) sensing MI, and (d) harvested power.}
\label{figure-MDDSCC-opPro}  
\vspace*{-1mm}
\end{figure}

The performance of Algorithm~\ref{MDDSCC:al1} in terms of communication rate, user association, sensing MI and harvested power are presented in Fig.~\ref{figure-MDDSCC-opPro}. As shown in Fig.~\ref{figure-MDDSCC-opPro}\,(a), the average rate per TS increases progressively with each outer-layer iteration, ultimately converging at the fourth iteration and attaining over 22\,bps/Hz. In addition, compared with the inner-layer iteration (i.e., steps~7-13 in Algorithm~\ref{MDDSCC:al1}), the outer-layer iteration adjusts time allocation (i.e., step 14 in Algorithm~\ref{MDDSCC:al1}) to bring a more significant improvement to the rate performance. Fig.~\ref{figure-MDDSCC-opPro}\,(b) illustrates the dynamic user association among the UAV and buoys. In particular, as the optimization process progresses, the UAV is prone to serve buoys 2 and 4 at the 21st and 31st TSs, respectively. Moreover, Fig.~\ref{figure-MDDSCC-opPro}\,(c) and Fig.~\ref{figure-MDDSCC-opPro}\,(d) show the achievable performance of MI and harvested power respectively after the 1st and 4th outer-layer iterations, where it can be observed that since the optimization objective is to maximize the uplink communication rates, the average MI/amount of harvested power per active TS\footnote{During the mission period, each buoy is only served by the UAV in certain TSs, and these TSs are denoted as `active TSs'.} deteriorate to some extent. But, owing to the constraints of \eqref{eqOPc3} and \eqref{eqOPc4}, the average MI per active TS of each buoy is still larger than the predefined threshold of 1\,bps/Hz, while the harvested power of around 2\,mW per active TS is sufficient to support the uplink communications.

\subsection{Impacts of System Structure and States}\label{S5.2}

\begin{figure}[t]
\vspace*{-1mm}
\centering
\includegraphics[width=0.8\linewidth]{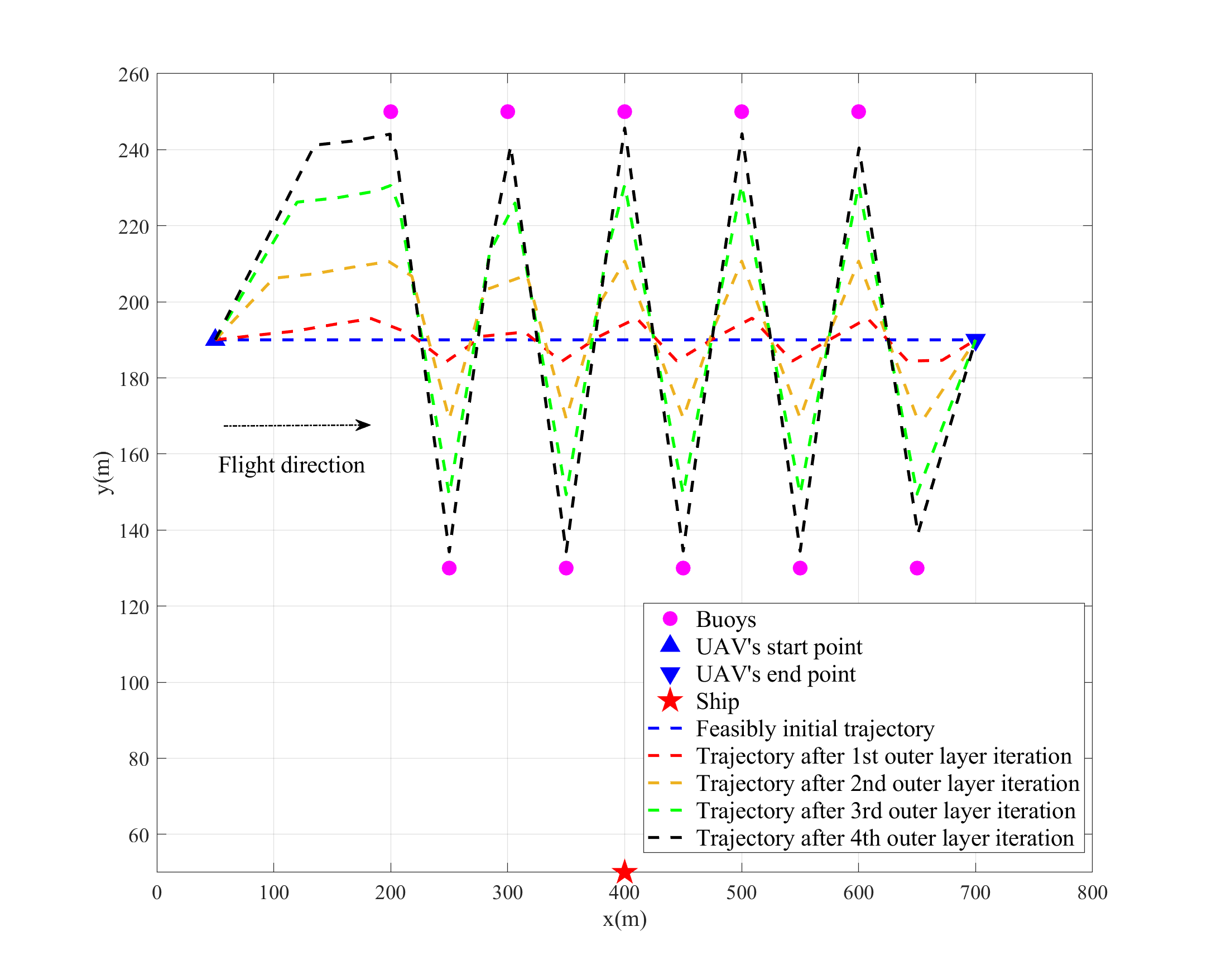}
\vspace*{-4mm}
\caption{\small Convergence behavior of the proposed optimization method for another buoy monitoring array.}
\label{figure-MDDSCC-TracOther} 
\vspace*{-3mm}
\end{figure}

To demonstrate the robustness of the proposed optimization method, we construct another buoy monitoring array as shown in Fig.~\ref{figure-MDDSCC-TracOther}, where 10 buoys constitute two parallel rows of linear monitoring array. As shown in Fig.~\ref{figure-MDDSCC-TracOther}, with the aid of the proposed method, the optimal UAV trajectory can be obtained after four outer-layer iterations.

\begin{figure}[!h]
\vspace*{-4mm}
\centering
\includegraphics[width=0.8\linewidth]{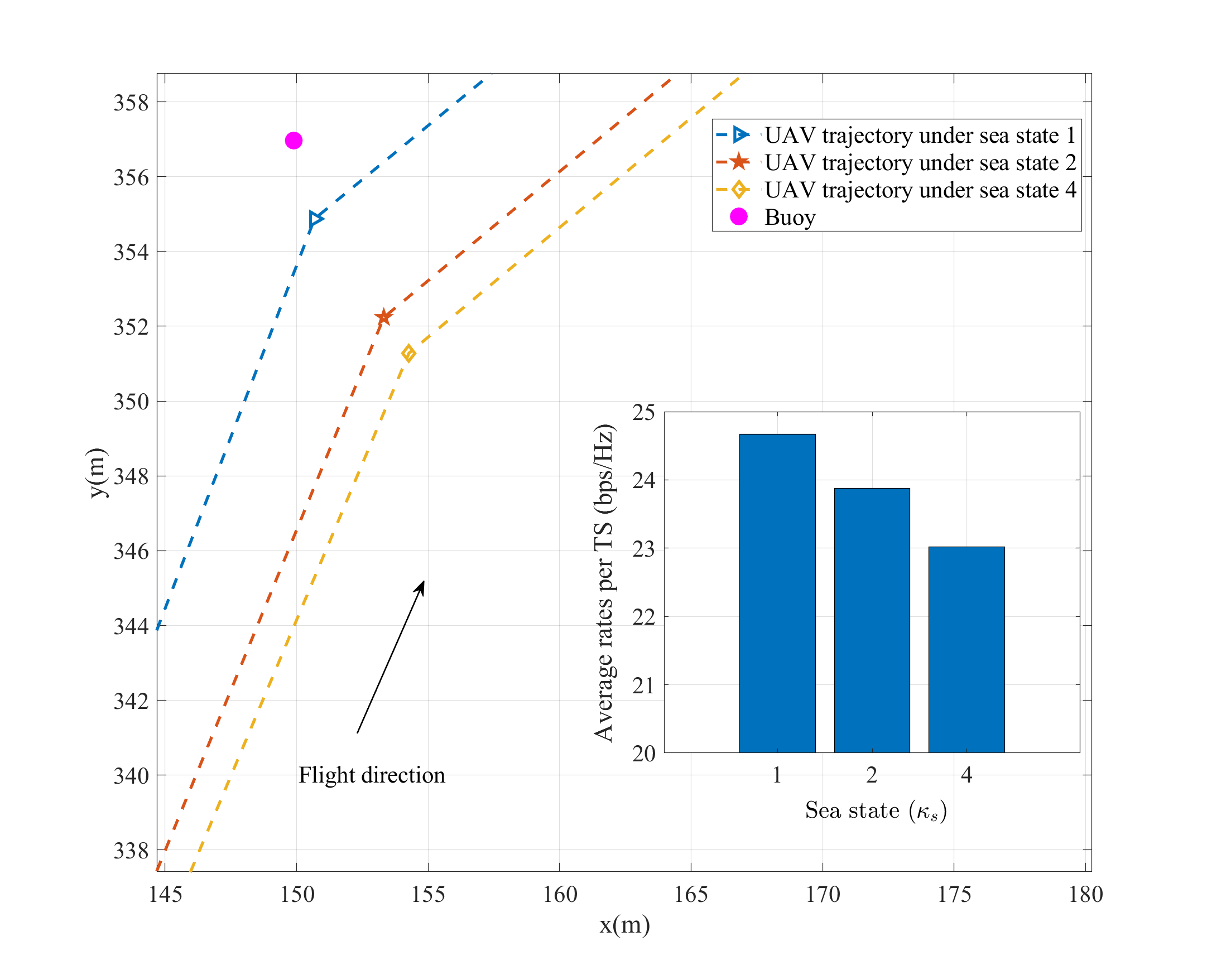}
\vspace*{-4mm}
\caption{\small The effect of sea clutter on the UAV trajectory and rate performance, where the sea states 1, 2, 4 are considered.}
\label{figure-MDDSCC-opSS}  
\vspace*{-1mm}
\end{figure} 

Next we analyze the effect of sea clutter on the performance of the proposed network. As shown in Fig.~\ref{figure-MDDSCC-opSS}, the UAV passes almost directly over the buoy under sea state 1, since sea clutter only comes from the grazing angle of around 80-90 degree, which is depicted in Fig.~\ref{figure-MDDSCC-sc}. When the sea state changes to 2, to avoid the sea clutter reflected from the grazing angle of around 60-90 degree, the UAV moves away from the buoy. When the sea state evolves from 2 to 4, the UAV trajectory does not change too much compared with that of sea state 2. This because with the increase of sea state, although the sea clutter arises from wider range of grazing angle, its power reduces a lot due to the irregular diffuse reflection caused by sea waves. Since the UAV progressively enlarges its stand-off distance from the buoy when the sea state increases from 1 to 4, the average rate per TS correspondingly reduces by 1.6\,bps/Hz as can be seen from the sub-figure of Fig.~\ref{figure-MDDSCC-opSS}. It can then be inferred that as the sea state becomes worse, the effect of sea clutter may not cause too much rate reduction.

\begin{figure}[!h]
\vspace*{-5mm}
\centering
\includegraphics[width=0.8\linewidth]{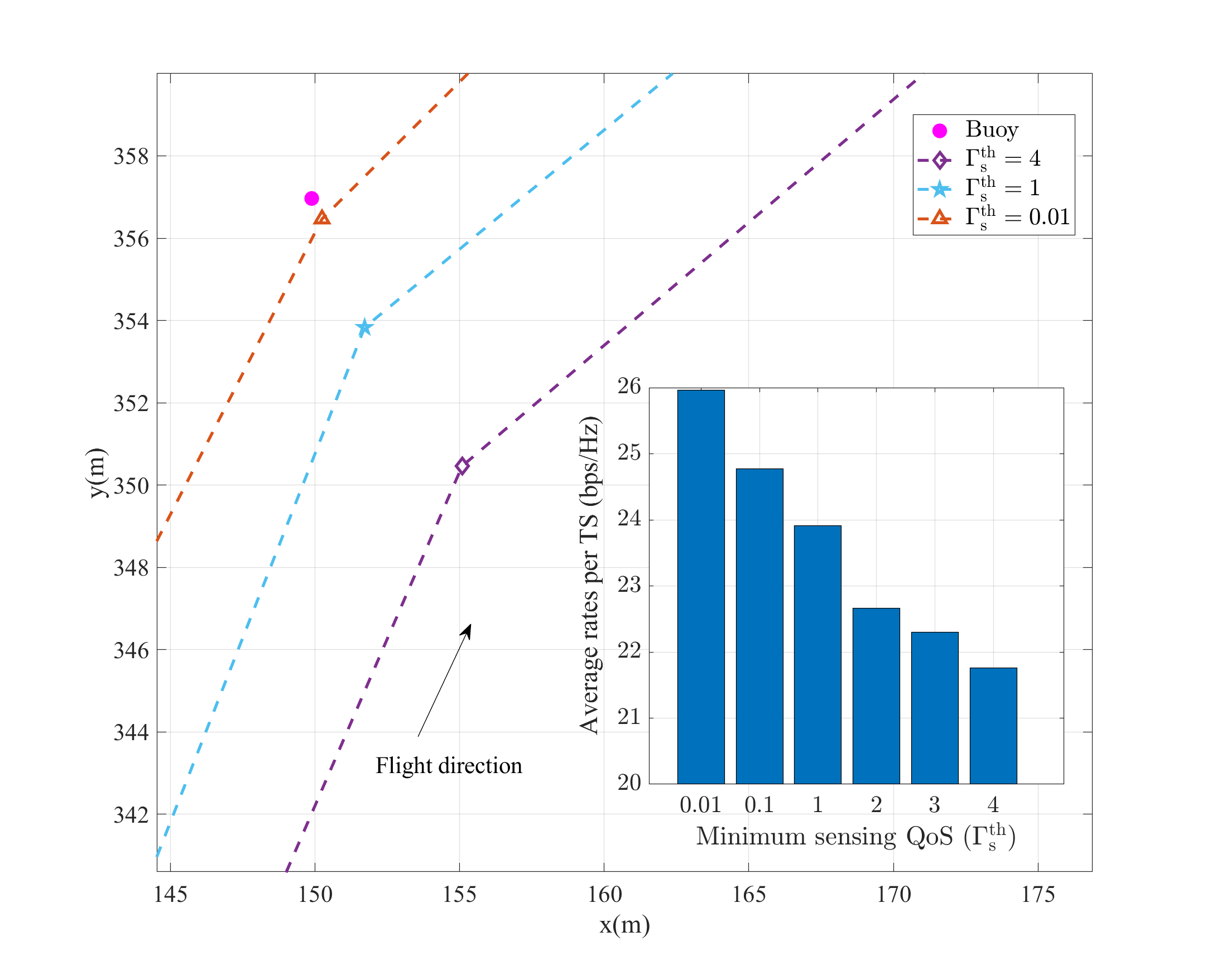}
\vspace*{-4mm}
\caption{\small The effect of minimum sensing QoS on the UAV trajectory and rate performance, where $\Gamma_{\text{s}}^{\text{th}}=0.01, 0.1, 1, 2, 3, 4$ and $\kappa_{\text{s}}=3$ are considered.}
\label{figure-MDDSCC-opMI} 
\vspace*{-1mm}
\end{figure} 

\begin{figure}[!b]
\vspace*{-5mm}
\centering
\includegraphics[width=0.8\linewidth]{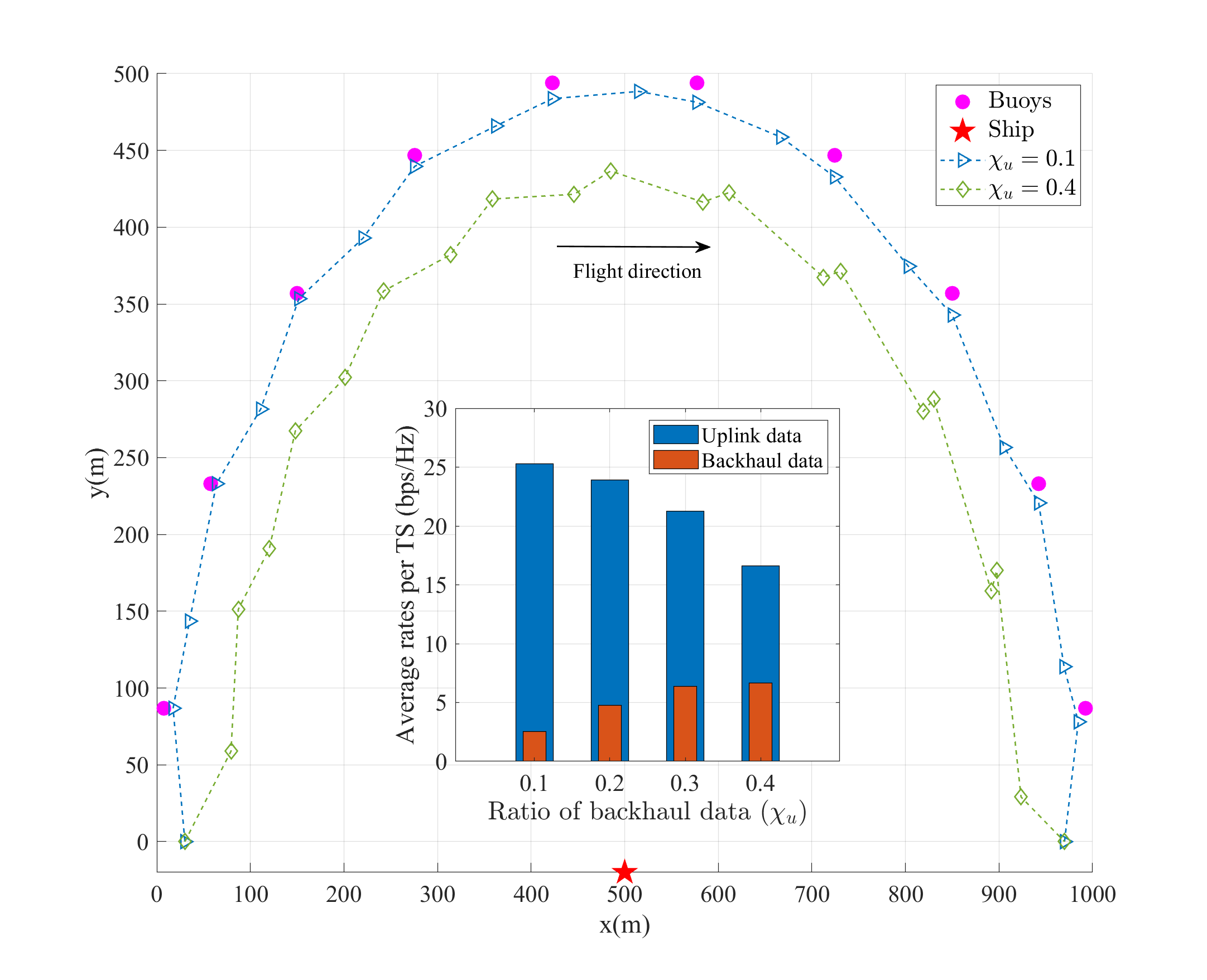}
\vspace*{-4mm}
\caption{\small The effect of backhaul transmission requirement on the UAV trajectory and rate performance, where $\chi_u\! =\! 0.1, 0.2, 0.3, 0.4$.}
\label{figure-MDDSCC-opBH} 
\vspace*{-1mm}
\end{figure} 

We then investigate the effect of the minimum sensing QoS on the performance of the proposed network. As expected, with the increase of $\Gamma_{\text{s}}^{\text{th}}$, the UAV allocates more time resource to satisfy the stringent requirement of sensing, leading to the average rate per TS reduction by 4.2\,bps/Hz from $\Gamma_{\text{s}}^{\text{th}}\! =\! 0.01$ to $\Gamma_{\text{s}}^{\text{th}}\! =\! 4$, which is shown in the sub-figure of Fig.~\ref{figure-MDDSCC-opMI}. It can also be seen from in Fig.~\ref{figure-MDDSCC-opMI} that increasing minimum sensing QoS causes the UAV moving away from the buoy because of the existing of sea clutter. In particular, to guarantee that all the buoys can be accurately sensed, the UAV keeps a safe distance from buoys so as to avoid being affected by sea clutter during receiving the desired sensing echo. Additionally, there is no doubt that the requirement of backhaul transmission may also highly impact the UAV flying strategy. 

The impact of backhaul transmission requirement on the performance of the proposed network is also studied in Fig.~\ref{figure-MDDSCC-opBH}. It can be seen that as the ratio of the backhaul data to the received uplink data increases, the UAV trajectory significantly changes and moves closer to the ship. Consequently, the amount of data uploaded to the ship increases. By contrast, the received uplink data reduces due to the larger path loss and shorter allocated time.

\begin{figure}[t]
\vspace*{-7mm}
\centering
\includegraphics[width=0.8\linewidth]{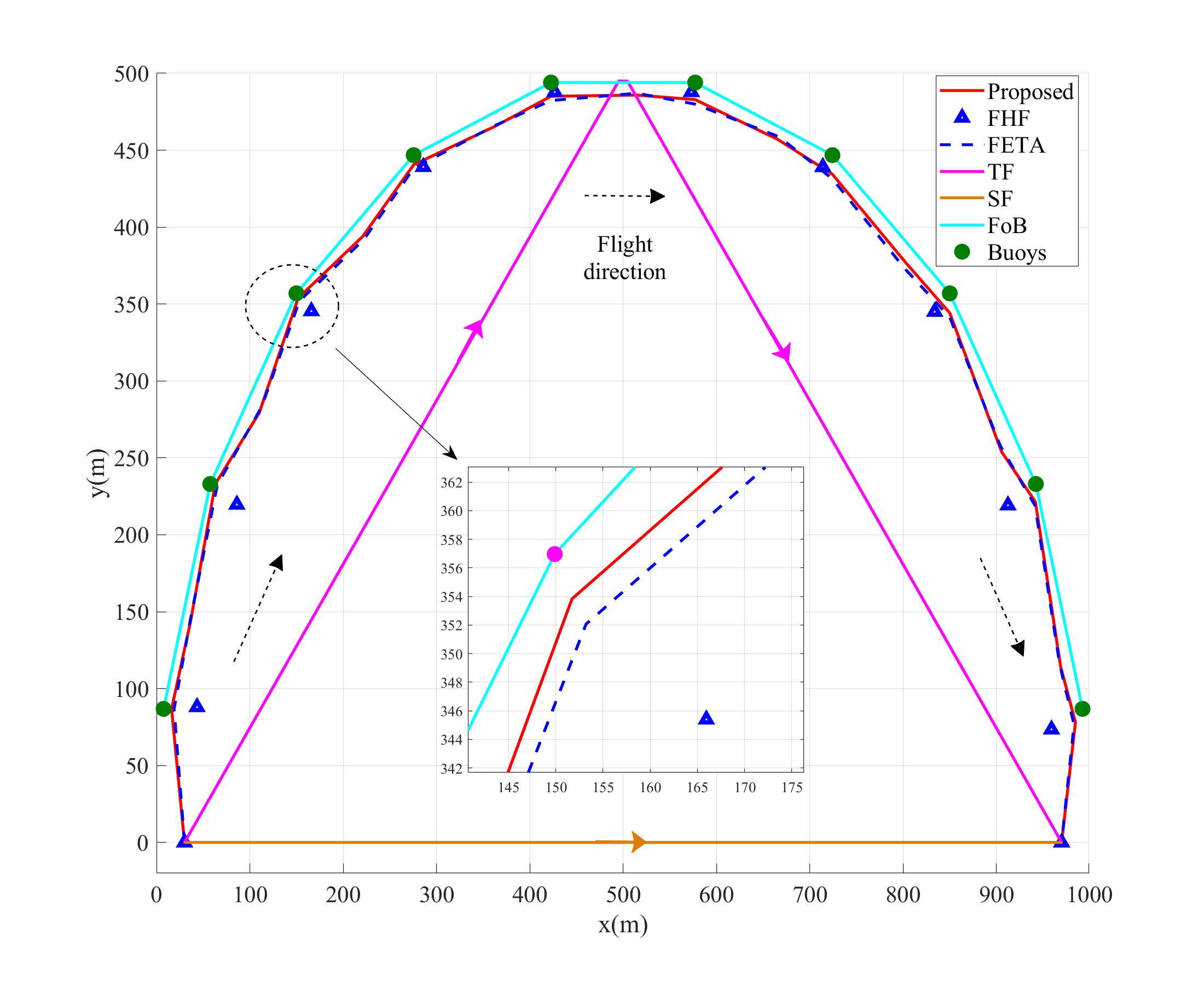}
\vspace*{-4mm}
\caption{\small UAV trajectory comparison between the proposed scheme and six benchmarks.}
\label{figure-MDDSCC-BenT}  
\vspace*{-1mm}
\end{figure} 

\begin{figure}[t]
\vspace*{-2mm}
\centering
\includegraphics[width=0.9\linewidth]{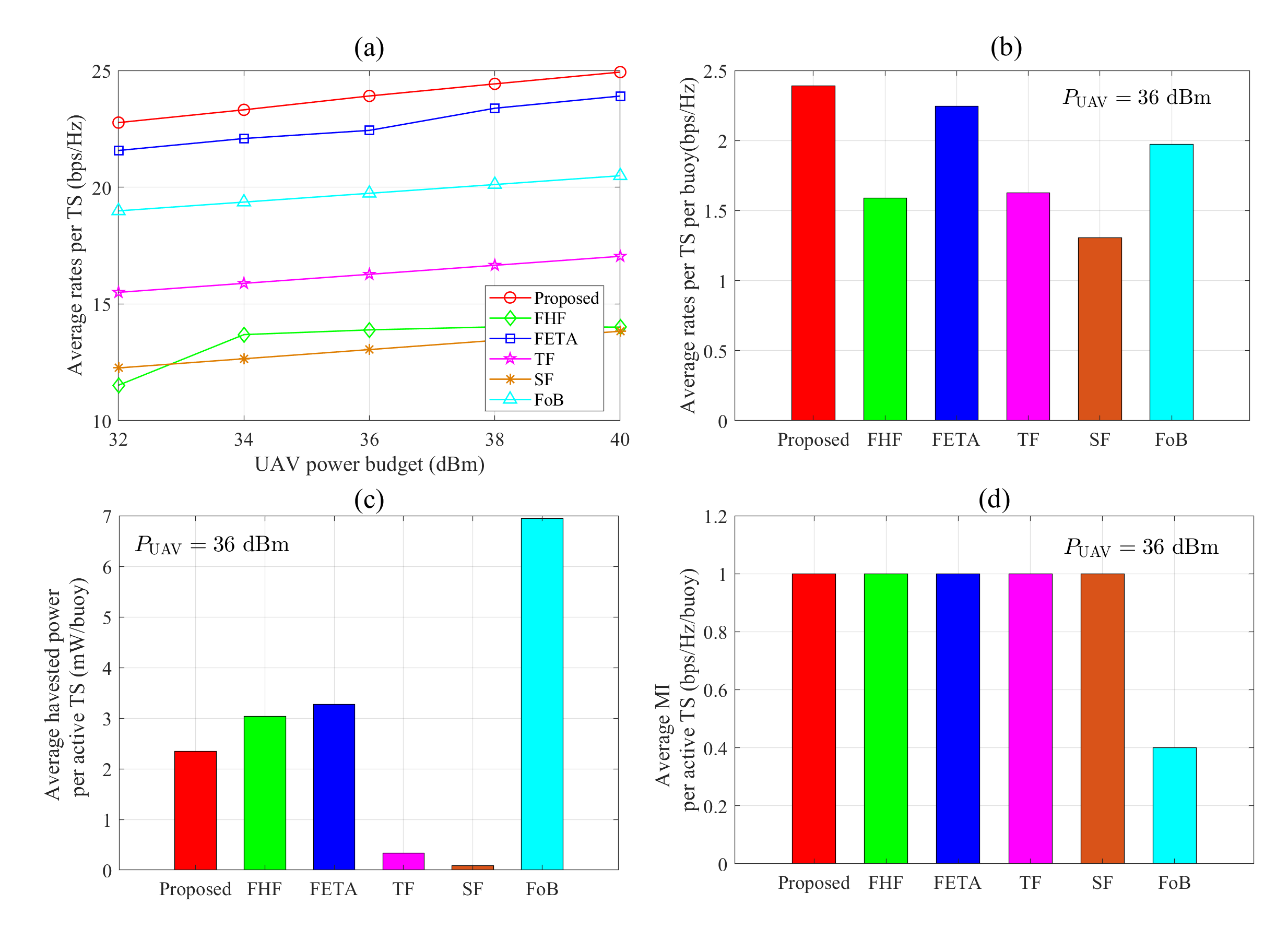}
\vspace*{-4mm}
\caption{\small The comparison between the proposed scheme and six benchmarks, in terms of (a)~average rate per TS, (b)~average rate per active TS per buoy, (c)~average harvested power per active TS per buoy and (d)~average MI per active TS per buoy. The results in (b)--(d) are obtained with $P_{\text{{UAV}}}\! =\! 36$\,dBm.}
\label{figure-MDDSCC-Ben}  
\vspace*{-3mm}
\end{figure} 

\subsection{Performance Comparison with Benchmarks}\label{S5.3}

Our scheme is compared with the following benchmarks for UAV-assisted maritime monitoring networks.

\noindent
$\bullet$~{\bf Fly Over Buoys (FoB)}: The UAV flies along the arc-shaped buoys monitoring array.

\noindent
$\bullet$~{\bf Fly-Hover-Fly (FHF)}: The UAV sequentially reaches the ten optimal hovering points, which are obtained by iteratively solving the transformed optimization problem (\underline{P2.1}) based on (P2.1).
\begin{subequations}\label{eq:MDDSCC:Opt2.1un} 
\begin{align}
  (\underline{\text{P}2.1}): & \max_{\underline{{\mathcal{S}}_4}} \sum_{n=1}^{N}\sum_{u=1}^{U}R^{\text{c}}_u[n], \label{Opt2.1un:eqOPob} \\
  \text{s.t.}~~ & \eqref{eqOPc3}-\eqref{eqOPc6}, \eqref{eqOPc9} \\
	& \sum_{u=2}^N \left\|\pmb{c}_{\text{UAV}}[n] - \pmb{c}_{\text{UAV}}[n-1]\right\| \leq V_{\text{max}}T,
\end{align}
\end{subequations} 
where $\underline{{\mathcal{S}}_4 }\! =\!\left\{\pmb{c}_{\text{UAV}}[\mathcal{N}],p_u[\mathcal{N}]\right\}$ with $\mathcal{N}\! =\! \left\{n_1,\dots,n_{10}\right\}$ is the set of ten hovering points and power scheduling factors. 

\noindent
$\bullet$~{\bf Triangle Fly (TF)}: The UAV flies along a predefined triangle path.

\noindent
$\bullet$~{\bf Straight Fly (SF)}: The UAV directly flies from the initial position to the destination.

\noindent
$\bullet$~{\bf Fly with Equal Time Allocation (FETA)}: The proposed UAV optimization strategy is adopted, while the equal time allocation stays unchanged during mission period. 

Unless otherwise stated, the optimization of the other related variables in each benchmark follows the way implemented in Algorithm~\ref{MDDSCC:al1}.
The optimal UAV trajectory of each scheme is depicted in Fig.~\ref{figure-MDDSCC-BenT}, and its corresponding performance are provided in Fig.~\ref{figure-MDDSCC-Ben}. 

As shown in Fig.~\ref{figure-MDDSCC-Ben}\,(a), our scheme outperforms all the benchmarks in terms of average rate per TS during mission period, and is able to achieve 25\,bps/Hz with 40\,dBm UAV's transmit power. FETA without optimizing time allocation lags behind the proposed scheme, but is still better than the other benchmarks, as it jointly considers the UAV trajectory, user association and power scheduling. In TF and SF, the predefined UAV paths diverge too much from the positions of buoys, leading to lower rates. As seen in Fig.~\ref{figure-MDDSCC-BenT} and Fig. \ref{figure-MDDSCC-Ben}\,(b), although the optimal hovering points of the UAV are close to buoys, FHF fails to optimize the path among any of two hovering points, leading to significant decrease of buoys' communication rates.

In terms of charging and sensing performance, FoB is capable of fulfilling the largest charging power as shown in Fig. \ref{figure-MDDSCC-Ben}\,(c). This is because the UAV flies along the distribution of buoys, and therefore has the shortest distance between buoys. On contrary, the UAV flying paths in TF and SF are far from buoys causing the unsatisfied performance of wireless charging. However, owing to the effect of sea clutter, FoB, where the UAV flies too close to buoys, cannot meet the requirement of minimum sensing MI of 1\,bps/Hz and only attains 0.4\,bps/Hz, while the other schemes are able to satisfy the sensing QoS. 

\section{Conclusions}\label{S6}

To improve the data collection in energy-limited buoy monitoring networks in open sea, we have proposed an ISCPB scheme with specifically-designed TDD-like frame structure, in which the UAV first localizes buoys using sensing signal, and then carries out wireless charging, data collection and backhaul transmission. Due to the effect of sea clutters, these multiple transmissions are tightly coupled in terms of time distribution, UAV trajectory, UAV-buoy association and power scheduling, which has motivated us to design a joint optimization method based on alternating optimization, quadratic transform and A-FOTE. Simulation results have demonstrated the effectiveness and good convergence behavior of the joint optimization method. In particular, our ISCPB scheme is capable of achieving an average data rate over 22\,bps/Hz per time slot relying only on the harvested power, which holds great potential to serve as a paramount data collection solution in open sea monitoring scenarios. 

\appendix

\subsection{Derivation of $R_u^{\text{c}}[n]$}\label{App:MDDSCC:Ruc}

Substituting $\tilde{\pmb{w}}_u[n]$ by $\pmb{a}_{\text{rx}}\left(\pmb{c}_{\text{UAV}}[n],\pmb{c}_u[n]\right)$ yields
\begin{align}\label{eqAp1} 
  R_u^{\text{c}}[n] =& \gamma_u^{\text{c}}[n] \log\Bigg(1 + \bigg(\frac{\alpha_u[n]p_u[n]G_{\text{gain}}^{\text{c}}\lambda^2}{N_0 4\pi^2 \breve{d}_u^2[n]}\bigg) \nonumber \\
  & \times \left|\pmb{a}_{\text{rx}}^{\text{H}}\left(\pmb{c}_{\text{UAV}}[n],\pmb{c}_u[n]\right)\pmb{a}_{\text{rx}}\left(\pmb{c}_{\text{UAV}}[n],\breve{\pmb{c}}_u[n]\right)\right|^2\Bigg) \nonumber \\
  & \overset{(a)}{=}\gamma_u^{\text{c}}[n] \log\Bigg(1 + \bigg(\frac{\alpha_u[n]p_u[n]G_{\text{gain}}^{\text{c}}\lambda^2}{N_0 4\pi^2 {d}_u^2[n]R^4}\bigg) \nonumber \\
  & \times \left|\sum_{r=0}^{R-1}e^{-j2\pi r \frac{e_x[n]}{2d_u^2[n]}}\sum_{r=0}^{R-1}e^{-j2\pi r \frac{e_y[n]}{2d_u^2[n]}}\right|^2\Bigg) \nonumber \\
  & \overset{(b)}{=}\gamma_u^{\text{c}}[n] \log\Bigg(1 + \left(\frac{\alpha_u[n]p_u[n]G_{\text{gain}}^{\text{c}}\lambda^2}{N_0 4\pi^2 {d}_u^2[n]R^4}\right) \nonumber \\
  & \times \left(\frac{\sin\big(R \hat{d}_u[n]e_x[n]\big)}{\sin\big(\hat{d}_u[n]e_x[n]\big)} \frac{\sin\big(R \hat{d}_u[n]e_y[n]\big)}{\sin\big(\hat{d}_u[n]e_y[n]\big)}\right)^2\Bigg) \nonumber \\
  & \overset{(c)}{=}\gamma_u^{\text{c}}[n] \log\Bigg(1 + \left(\frac{\alpha_u[n]p_u[n]G_{\text{gain}}^{\text{c}}\lambda^2}{N_0 4\pi^2 {d}_u^2[n]}\right)\Bigg) ,  
\end{align}
where $(a)$ is obtained due to the fact that under the good sea condition, the small wave-induced displacement (i.e., $e_x[n]$, $e_y[n]$ and $e_z[n]$) is much shorter than the distance between buoy $u$ and the UAV, i.e., $e_x[n],e_y[n],e_z[n]\! \ll\! d_u[n]$, and hence $d_u[n]\! \approx\! \breve{d}_u[n]$. Step $(b)$ is obtained using $\sum_{n=0}^{N-1}{e^{-j2\pi nx}}\! =\! \frac{\sin(\pi N x)}{\sin(\pi x)}e^{-j\pi x(N-1)}$, where $\hat{d}_u[n]\! =\! \frac{\pi}{2d_u[n]}$, and step $(c)$ is derived using $\lim_{x\rightarrow 0} \frac{\sin ax}{x}\! =\! a$.



\end{document}